\documentclass[reprint,secnumarabic,amssymb, nobibnotes, aps, prb]{revtex4-1}

\usepackage{graphicx}
\usepackage{graphics}
\usepackage[percent]{overpic}
\graphicspath{ {images/} }
\usepackage{amsmath}
\usepackage{mathtools}

\usepackage[utf8]{inputenc}
\usepackage{amsfonts}
\usepackage{amssymb}
\usepackage{float}
\usepackage{comment}
\usepackage{color,soul}
\usepackage{textcomp}
\usepackage{subcaption}
\usepackage{hyperref}
\usepackage[dvipsnames]{xcolor}
\newcommand{\vk}{{\bf k}}

\newcommand{\cH}{\mathcal{H}}

\newcommand{\id}{\,\mathrm{d}}
\newcommand{\intk}{\,\int  \frac{\id ^2 k}{(2\pi)^2}}

\newcommand{\intkprfbz}{\,\int\frac{\id ^2 k'}{(2\pi)^2}}

\DeclareMathOperator{\sech}{sech}

\begin{document}

%%%%%%%%%%%%%%%%%%%%%%%%%%%%%%%%%%%%%%%%%%%%%%%%%%%%%%%%%%%%%%%%%%%%%%%%%%%%%%%%%%%%%%%%%%%%%%%%%%%%%%%%
\title{Thermal Transport in 2D Nematic Superconductors}%

\author{Sourav Sen Choudhury, Sean Peterson, Yves Idzerda}%
%\email[REVTeX Support: ]{revtex@aps.org}
\affiliation{Department of Physics, Montana State University, Bozeman, Montana 59717, USA}
\date{\today}
\begin{abstract}
        We study the thermal transport in a two-dimensional system with coexisting superconducting (SC) and nematic orders. We analyze the nature of the coexistence phase in a tight-binding square lattice where the nematic state is modelled as a $d$-wave Pomeranchuk type instability and the feedback of the symmetry breaking nematic state on the SC order is accounted for by mixing of the $s$, $d$ paring interaction. The electronic thermal conductivity is computed within the framework of Boltzmann kinetic theory where the impurity scattering collision is treated in the both the Born and Unitary limits. We present qualitative, analytical, and numerical results that show that the heat transport properties of SC states emerging from a nematic background are quite distinct and depend on the degree of anisotropy of the SC gap induced by nematicity. We describe the influence of the Fermi surface topology, the van Hove singularities, and the presence or absence of zero energy excitations in the coexistence phase on the the low temperature behaviour of the thermal conductivity. Our main conclusion is that the interplay of nematic and SC orders has visible signatures in the thermal transport which can be used to infer SC gap structure in the coexistence phase.
\end{abstract}
\maketitle
%\tableofcontents

%%%%%%%%%%%%%%%%%%%%%%%%%%%%%%%%%%%%%%%%%%%%%%%%%%%%%%%%%%%%%%%%%%%%%%%%%%%%%%%%%%%%%%%%%%%%%%%%%%%%%%%%
\section{Introduction}

Low temperature transport properties of normal metals  are primarily determined by the scattering of electrons by impurities. For heat transport, the linear  $T$ dependence of the thermal conductivity, $\kappa_n(T)$, can be explained using semi-classical transport theory based on the Boltzmann kinetic equation\cite{ziman} which has also been used to explain heat transport properties of conventional superconductors \cite{bardeen}. The advent of unconventional superconductors  like heavy fermions \cite{hfrev}, cuprates  \cite{harlin, tsu, tail, pair} and iron-based superconductors \cite{Wen, stewart, chubukov}, lead to new questions regarding the low temperature transport properties of such systems since the unconventional superconductors significantly differ from the uniformly gapped conventional superconductors and their gap structure may contain nodal points (i.e points on the Fermi surface (FS) where the superconducting gap is zero). The small energy gap surrounding the nodal points allows quasiparticles to be easily excited and hence these nodal quasiparticles dominate the heat transport properties at low temperatures. Thermal transport in unconventional superconductors has been previously studied theoretically by various authors with different levels of sophistication \cite{arfi, Hirsch,scharn,mon, durst, graf}, and thermal conductivity measurements are a useful probe of the gap structure of unconventional superconductors.\cite{Matsuda, Shak}

Unconventional superconductors possess complex phase diagrams with multiple broken symmetry phases coexisting with superconductivity. Often these multiple phases appear at similar ordering temperatures when material properties (like dopant concentration) are varied over wide ranges. While it is fairly common for unconventional superconductors to have proximate magnetic and superconducting orders \cite{lake,mathur,Badoux2016,kim,ley}, only in recent years have nematic states been reported for both iron-based superconductors \cite{chuang2010,BOHMER201690, chu2012} as well as cuprates \cite{Nakata2021,ando2002,hinkov2008,Sato2017, cyr2015,Wu2017}. (Here nematic order means electronic nematicity, where the electronic state has the same translational symmetry as the underlying crystal, but a lower rotational symmetry.)  Studies on the origin of the nemetic state \cite{Fernandes2014} argue that in iron-based superconductors, nematic order is driven by either spin fluctuations \cite{HU2012215,Fernandes2013-2} (in the case of pnictides) or orbital fluctuations \cite{Bohmer2013, yamakawa2016, fanfarillo2018} (in the case of chalcogenides). For cuprates it has been proposed that the nematicity arises from fluctuations of stripe order \cite{kivelson2003, fradkin2010} or from the instability of the Fermi surface (Pomeranchuk instability). \cite{yamase2006,oganesyan2001,kao2005,halboth2000} 

Regardless of the origin of the nematic state, the influence of nematicity on the emerging superconducting state can change the character of the superconducting order from $s$-wave to $d$-wave pairing \cite{Fernandes2013}. Additionally, since the anisotropy of the superconducting state correlates with the Fermi surface deformation of the nematic state, the competition or cooperation between the SC and nematic orders is found to depend on the nematic distortion of the Fermi surface relative to the anisotropy of the superconducting gap function. \cite{chen2020}

Nematic superconductors themselves may display interesting thermal transport behavior. For instance, the nematic to isotropic quantum phase transition deep within the $d$-wave superconducting phase of a two-dimensional tetragonal crystal are predicted, within the framework of the Boltzmann equation, to display a logarithmic enhancement of the thermal conductivity at the nematic critical point \cite{fritz}. Other theoretical studies, performed using the quasi-classical formalism, show that the oscillations of the thermal conductivity in multi-band superconductors with an anisotropic gap under a rotating magnetic field, change sign at low temperatures and fields and can be used to distinguish between nodes and minima in the energy gap of iron-based superconductors.\cite{vorontsov2010, chubukov2010}

Recent experimental studies have examined the structure of the SC gap in iron-based nematic superconductors. Using specific heat measurements, it was found that the electronic specific heat was linear in $T$ for $T<T_c$, indicating the presence of line nodes \cite{hardy2019} while  angle-resolved photo-emission spectroscopy (ARPES) \cite{kush2020} observed spontaneous breaking of the rotational symmetry of the SC gap amplitude as well as the unidirectional distortion of the Fermi pockets. (It should be noted that this latter study indicated that in the compound LiFeAs, nematicity could occur below $T_c$ and speculated that superconducting state develops a spontaneous nematic order at $T_c$.)

The gap structure of nematic superconductors have also been probed by thermal conductivity experiments \cite{dong2009, hope2016} demonstrating that in the $T\rightarrow 0$ limit, the residual linear term, $\kappa(T)/T$, is extremely small, indicating nodeless superconductivity in FeSe. Finally in the case of cuprate superconductors \cite{ando2002, hinkov2008} and strontium ruthenate materials \cite{borzi2007}, transport measurements  show large strongly temperature-dependent anisotropies in these otherwise isotropic electronic systems.

Motivated by these experimental studies and in complement to previous theoretical studies, this paper investigates the thermal transport properties of a nematic system, where the superconducting phase arises out of a nematic background (i.e. the onset of SC order occurs at a lower temperature than the nematic order). To treat the nematic and SC orders on equal footing, we introduce a mean field Hamiltonian and determine how the interaction between these coexisting phases impacts the heat transport properties of the system. For our transport calculations we use the quasiparticle Boltzmann equation (which is physically more transparent than calculations based on the Green's function or quasi-classical methods), and calculate the thermal conductivity for the case where the dominant scattering process of quasiparticles is by nonmagnetic impurities.  Within Boltzmann theory, we only consider the case of small phase shifts (i.e. the Born approximation) and phase shifts close to $\pi/2$ (i.e the unitary limit). The quasiparticle Boltzmann approach fails at low temperatures when low-energy quasiparticles cannot be well-established due to impurity broadening. In the following we assume that a quasiparticle description applies. \cite{kim2008} 

The organization of the paper is as follows. In sections \ref{sec:model}.\ref{sec:H}, we discuss the model Hamiltonian and the formalism we have employed.  The self-consistent approach to determining co-existing nematic and SC order parameters is presented in section \ref{sec:model}.\ref{sec:selfconsistency} and the kinetic formalism  is described in section \ref{sec:model}.\ref{sec:kinetic}. Numerical results for heat conductivity are discussed in section \ref{sec:results}. Section \ref{sec:conclusion} is a brief conclusion.

%%%%%%%%%%%%%%%%%%%%%%%%%%%%%%%%%%%%%%%%%%%%%%%%%%%%%%%%%%%%%%%%%%%%%%%%%%%%%%%%%%%%%%%%%%%
\section{Model and Formalism}
\label{sec:model}

\subsection{\label{sec:H} Hamiltonian}

\begin{figure}[t!]
\begin{centering}
\includegraphics[width=8.5cm]{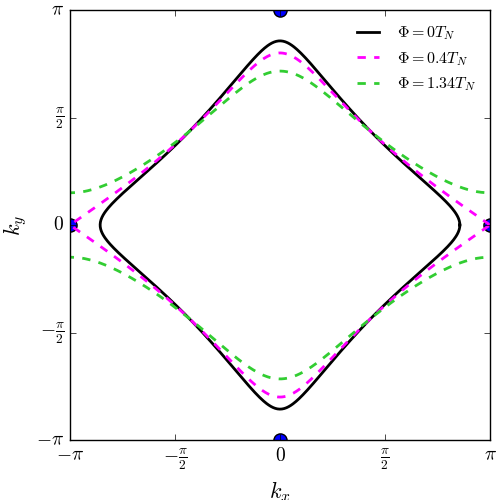}
\caption{Evolution of the Fermi surface shape under nematic distortion at different temperatures. Closed FS ($\mu = -4.8 T_N$) at $T > T_N$ (black curve), at $T = 0.97 T_N$ (magenta curve), and $T = 0$ (green curve). The blue dots indicate the locations of the saddle points in the band structure that lead to van Hove singularities in the bare DOS (see FIG.~\ref{fig:dos_vHs}). It can be seen that the magenta curves pass through the saddle points at $(\pi,0)$. The band parameters are $t_1 = 6 T_N$ and $t_2 = -T_N$. } 
\label{fig:FS}
\end{centering}
\end{figure}

For our model we consider a 2D system with a single band with an inversion symmetric dispersion $\xi_\vk (=\xi_{-\vk})$ given by
%\blue{In this model a single-band 2D system was considered with an inversion symmetric dispersion ($\epsilon(\vk) = \epsilon(-\vk)$) and chemical potential $\mu$}
\begin{equation}
 H_0 =\sum\limits_{\vk ,\sigma = \pm1} \xi_{\vk }c_{\vk \sigma}^{\dagger}c_{\vk \sigma},   
\end{equation}
where 
$$ 
\xi_\vk= -2\big[t_1(\cos k_x + \cos k_y) + 2 t_2\cos k_x \cos k_y\big] - \mu .
$$
This describes the nearest neighbor and next-nearest neighbor hopping on a 2D square lattice with lattice spacing $a=1$.
  The nematic state is  modelled through an additional mean field Hamiltonian \cite{yamase2005} 
  \begin{align}
  H_{nem} &= \sum\limits_{\vk ,\sigma } \Phi f_{\vk} c_{\vk \sigma}^{\dagger}c_{\mathbf{k}\sigma}\nonumber
  \\
  \Phi &=-V_{nem}\sum\limits_{\vk} f_{\vk}\langle c_{\vk}^{\dagger}c_{\vk }\rangle
  \label{ham_nem}
  \end{align}
  where $\Phi$ is the nematic order parameter and $f_{\vk}=(\cos k_x - \cos k_y$). This additional term causes a deformation of the Fermi surface (FS) which elongates it along the $k_x$-axis and shrinks it along the $k_y$-axis as is illustrated in FIG.~\ref{fig:FS}. Thus in the nematic state (when $\Phi\neq0$) the deformed FS  does not have the same point group symmetry of the underlying 2D lattice and can capture the effect of symmetry-breaking FS deformations on the SC state\cite{chen2020}. In this paper, only the case where the nematic transition temperature is greater than the superconducting critical temperature ($T_N > T_c$) is considered, (i.e. superconductivity arises inside the nematic state). 
  
  The effect of the symmetry-broken nematic state on the development of the SC order can be accounted for by using a SC order parameter of the form\cite{chen2020}
  
  $$
  \Delta_{\vk}=\Delta \mathcal{Y_{\vk}}
  $$  
  where $\mathcal{Y}_{\vk} = (1 + rf_{\vk})/\sqrt{1+r^2}$ ($\mathcal{Y}_\vk$ is normalized by $\sqrt{1+r^2}$ to ensure that $\int\frac{d^2k}{(2\pi)^2}|\mathcal{Y}_\vk|^2 = 1$) and $\Delta_\vk=\Delta_{-\vk}$. Here $r$ is a phenomenological anisotropy parameter and is a measure of the degree of anisotropy caused by the coexisting nematic order.  (The anisotropy parameter $r$ is proportional $\Phi$ and when $\Phi$ is zero, the SC interaction reduces to pure $s$-wave.) This form of the order parameter encapsulates the mixing of the $s$ and $d$-wave components induced by nematicity (it is assumed that superconductivity only exists in the spin singlet channel). While $r \propto \Phi$, it should be noted that it also depends on details of the electronic structure \cite{chen2020} that are beyond the scope of this work (hence $r$ is treated as a phenomenological parameter). In the nematic state, $r\neq 0$ and can be either positive or negative. 
  
\begin{figure}[t!]
\begin{centering}
\includegraphics[width=8.5cm]{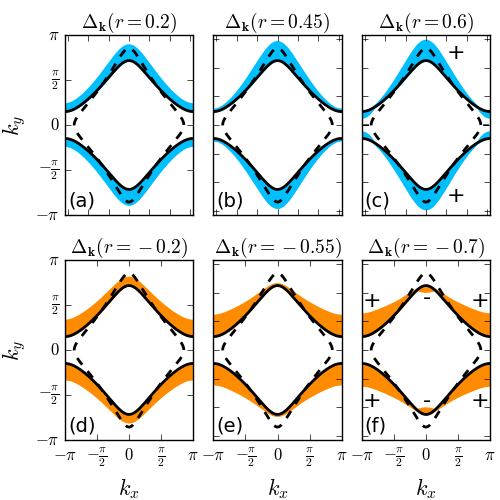}
\caption{The superconducting gap for different values of the anisotropy parameter, $r$, along the FS deformed by the nematic order (dotted line indicates original FS, solid line indicates deformed FS). For positive values of $r$, the direction of the the SC gap maximum (cyan) is anti-aligned with the FS elongation. For negative values of $r$, the direction of the the SC gap maximum (orange) is aligned with the FS elongation. The parameters used for the illustration are $\mu = -4.8 T_N$, $t_1 = 6 T_N$, $t_2 = - T_N$, $\Delta = 0.2 T_N$, and $\Phi = 1.34 T_N$}.
\label{fig:compVcoop}
\end{centering}
\end{figure}

  In FIG.~\ref{fig:compVcoop}, the non-uniform SC gap is shown as a colored band bordering the deformed FS for different values of the anisotropy parameter, $r$. As shown in the figure, the direction of the SC gap maximum relative to the direction FS elongation (induced by the nematic order) depends on whether $r$ is positive or negative. Thus, the superconducting part of the mean field Hamiltonian can be written as 
\begin{align}
H_{SC} &= \frac12 \sum\limits_{\vk ,\sigma} \sigma \Delta \mathcal{Y}_{\vk} \left(c_{\vk \sigma}^{\dagger}c_{-\vk  -\sigma}^{\dagger} + h.c.\right) \nonumber 
\\
\Delta &=-V_{sc}\sum\limits_{\vk }\mathcal{Y}_{\mathbf{k}}\langle c^{\dagger}_{\mathbf{-k},\downarrow}c^{\dagger}_{\vk ,\uparrow}\rangle
\label{ham_sc}
\end{align}
  
and the full mean field Hamiltonian for intertwined nematic and superconducting orders given by
$$
H = H_0 + H_{nem} +H_{SC}
$$
can be recast into a matrix form for particular spin orientations $\sigma=\pm 1 (\uparrow,\downarrow)$
\begin{eqnarray}
\begin{aligned}
&H^{(\sigma)}=\frac{1}{2}\sum\limits_{\vk} \hat{\Psi}_{\vk,\sigma}^{\dagger} \hat{\cH}^{(\sigma)}_{\vk} \hat{\Psi}_{\vk,\sigma}
\\
&\hat{\cH}^{(\sigma)}_{\vk} =
\begin{pmatrix}
\xi_\vk + \Phi f_\vk  & \sigma \Delta_{\vk }  \\
\sigma \Delta_{\vk } & - \xi_{\vk } - \Phi f_\vk\\
\end{pmatrix}
\label{ham full}
\end{aligned} 
\end{eqnarray}
where $\hat{\Psi}_{\vk,\sigma}^{\dagger} = \begin{pmatrix}c_{\vk \sigma}^{\dagger},c_{-\vk -\sigma} \end{pmatrix}$ is the Nambu vector. The leading factor of $1/2$ is from the particle-hole doubling of the bands in superconductivity. The eigenvalues of $\hat{\mathcal{H}}^{(\sigma)}_{\vk}$ give the quasiparticle energies, $\pm E_\vk$, where 
\begin{equation}
E_\vk=\sqrt{\bigg(\xi_\vk +\Phi f_\vk\bigg)^2 +\Delta^2_\vk}.
\end{equation}
As noted earlier, the nature of the spectrum critically depends on the value of the anisotropy parameter, $r$. When $r>0$, the spectrum has nodes (i.e. points on the nematic FS for which $E_\vk=0$) only if the parameter $r$ exceeds a critical value $r > r_c^+$ where  $r_c^+ = -\frac{2t_1t_2 + t_2\Phi - 4t_2^2}{4t_2^2 + t_2\mu - 4t_1t_2}$. When $r<0$, the spectrum has nodes only if the parameter $r$ is below a critical value $r < r_c^-$ where $r_c^- = -\frac{2t_1t_2 + t_2\Phi + 4t_2^2}{4t_2^2 + t_2\mu + 4t_1t_2}$. These critical values, $r_c^{\pm}$, can be determined from the condition $E_{\vk} = 0$, which occurs only when $\Tilde{\xi}_{\vk}\equiv \xi_\vk +\Phi f_\vk$ and $\Delta_\vk$ simultaneously vanish. 

To find the location of the nodes we set
\begin{small}
\begin{equation}
\Tilde{\xi}_{\vk} = 0 \Rightarrow k_y^{*} =  \cos^{-1}\bigg(-\frac{\mu + 2t_1 \cos k_x - \Phi \cos k_x}{2t_1 + 4t_2 \cos k_x + \Phi}\bigg)
\end{equation}
\end{small}which gives us the $k_y$ coordinates of all points along the nematically deformed FS on the upper half of the BZ as a function of $k_x$. To find the locations of the nodes on the deformed FS, we set
\begin{small}
\begin{align}
\begin{split}
    \Delta_{(k_x,k_y^*)} &= 0 \Rightarrow k_x^{\pm} = \cos^{-1}\bigg(\frac{-t_2-rt_1 \pm p}{2rt_2}\bigg) \\
    p &= \sqrt{t_2^2+r^2t_1^2-rt_2\Phi-r^2\mu t_2}
    \label{kx_node}
\end{split}
\end{align}
\end{small}

\begin{figure}[t]
\begin{centering}
\includegraphics[width=7.8cm]{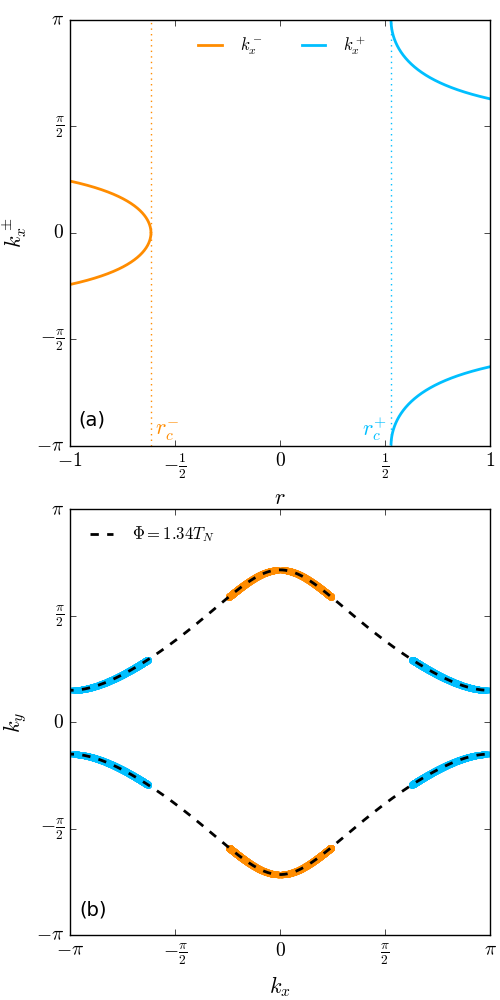}
\caption{(a) Equation (\ref{kx_node}) has solutions $k_x^{+}$ only when $r > r_c^+$ (cyan curve) and $k_x^{-}$ only when $r < r_c^-$ (orange curve). With the parameters $\mu = -4.8 T_N$, $t_1 = 6 T_N$, $t_2 = - T_N$, and $\Phi = 1.34 T_N$, the critical $r$-values are $r_c^+ \approx 0.52866$ and $r_c^- \approx -0.61447$. The dotted vertical lines emphasize that there are no solutions to equation (\ref{kx_node}) when $r$ is in the range $r_c^- < r < r_c^+$. (b) Range of locations $(\pm k_x^+, \pm k_y^*)$ of the nodes when $r_c^+ < r < 1$ (shaded in cyan) and $(\pm k_x^-, \pm k_y^*)$ when $-1 < r < r_c^-$ (shaded in orange). A particular $r$-value only corresponds to point nodes located either in the cyan regions or in the orange regions.}
\label{fig:nodeLocations}
\end{centering}
\end{figure}

In FIG.~\ref{fig:nodeLocations}(a) we display $k_x^{\pm}$ as a function of $r$, which identifies the critical values $r_c^{\pm}$ and shows that nodes only exist at $k_x^+$ when $r$ is positive and at $k_x^-$ when $r$ is negative. The location of these nodes depends on the value of the parameter $r$. FIG.~\ref{fig:nodeLocations}(b) shows the range of locations of the point nodes on the deformed FS as $r$ takes values in the range $r_c^+ < r < 1$ (region shaded in cyan) and $-1 < r < r_c^-$ (region shaded in orange). It should be emphasized that at any given $r$-value only a single point node exists in each quadrant of the BZ, the shaded regions only represent the range of locations.

\begin{figure}
\begin{centering}
\includegraphics[width=7.9cm]{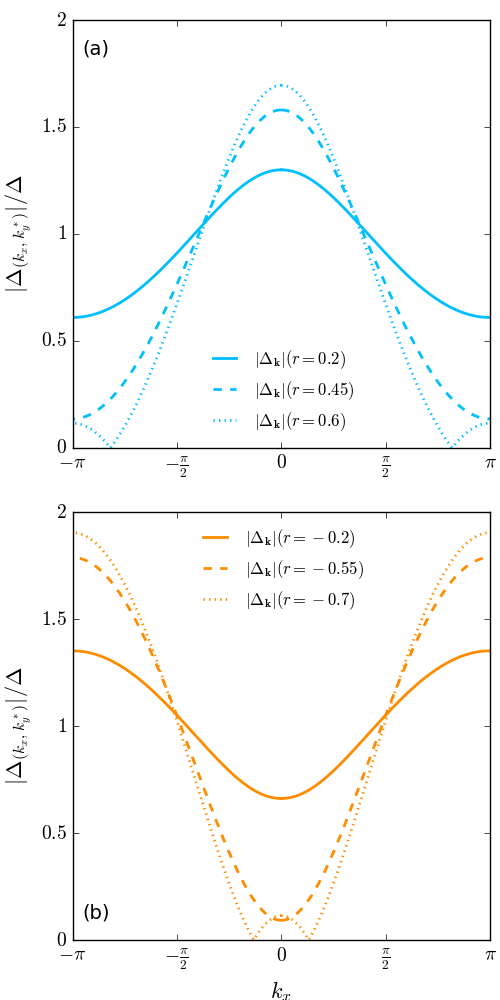}
\caption{Gap amplitude $|\Delta_\vk| $ in the coexistence phase along the nematically deformed FS ($\Tilde{\xi}_\vk = 0$) at $T = 0$ with parameters $\mu = -4.8 T_N$, $t_1 = 6 T_N$, $t_2 = - T_N$, $r_c^+ \approx 0.52866$, and $r_c^- \approx -0.61447$. (a) Low-energy excitations when $0 < r < r_c^+$ occur at $(\pm \pi,\pm k_y^*)$ in the BZ before the appearance of nodes. When $r > r_c^+$ secondary local maxima of the SC gap amplitude appear at $(\pm \pi,\pm k_y^*)$. (b) Low-energy excitations when $r_c^- < r < 0$ occur at $(0,\pm k_y^*)$ in the BZ before the appearance of nodes. When $r < r_c^-$ secondary local maxima of the SC gap amplitude appear at $(0,\pm k_y^*)$.}
\label{fig:energyGap}
\end{centering}
\end{figure}

In FIG.~\ref{fig:energyGap} we plot the $|\Delta_\vk|$ along the deformed FS. We see that for $0<r<r_c^+$, $|\Delta_\vk|$ has minima at $(\pm \pi, \pm k_y^*)$, whereas for $r_c^-<r<0$, the minima occur at $(0,\pm k_y^*)$. Therefore, these also indicate the locations of the excitations with the lowest energies. However, once the nodes form (i.e. for $r > r_c^+$ or $r < r_c^-$),  $|\Delta_\vk|$ has a secondary local maxima at these same locations in the BZ. The location of the low-energy excitations (before the formation of nodes) and the appearance of these secondary maxima of the gap amplitude (after the formation of nodes) have a significant effect on the heat transport propreties of the system. (see Section \ref{sec:results}).

While the presence of the nodes plays a dominant role in determining the transport properties in the coexistence phase at low temperatures (as will be discussed later in Section \ref{sec:results}), the existence of van Hove singularities is an important feature that influences transport properties for $T > T_c$ when the system is in the purely nematic phase. The dispersion relation in the nematic phase ($\tilde{\xi}_\vk = \xi_\vk + \Phi f_\vk$) has saddle points ($|\nabla_\vk \Tilde{\xi}_\vk| = 0$) close to the FS at $(k_x,k_y) = (\pi,0)$ and $(0,\pi)$, which can be seen in Fig.~\ref{fig:FS} as the blue points. These saddle points cause van Hove singularities to occur in the bare density of states at energies $\Tilde{\xi}_\vk^{vH} = - \mu + 4t_2 \pm 2\Phi$. Furthermore, as can be seen in FIG.~\ref{fig:FS}, the nematic FS passes through these saddle points when the closed FS transitions to an open FS along the $\hat{k}_x$-axis. 
\begin{figure}
\begin{centering}
\includegraphics[width=8.5cm]{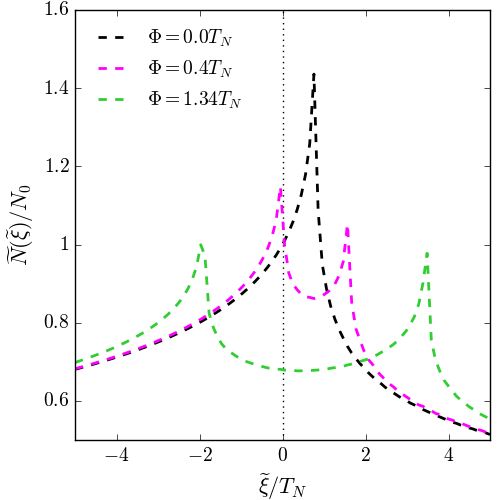}
\caption{Evolution of the bare DOS under the nematic order at different temperatures for a closed ($\mu = -4.8 T_N$) FS. The DOS at $T > T_N$ and at $T = 0$ are given by the black and green curves in. The van Hove singularities in the bare DOS cross the Fermi level at $T = 0.97 T_N$ (magenta curve). This occurs when the deformed FS passes through the saddle point located at $(\pi,0)$ as seen in FIG.~\ref{fig:FS} (see text for details).}
\label{fig:dos_vHs}
\end{centering}
\end{figure}

In the absence of nematicity, the saddle points at $(k_x,k_y) = (\pi, 0)$ and $(0,\pi)$ lead to van Hove singularities in the bare DOS at the same energy\cite{yamase2005} ($\xi_\vk^{vH} = -\mu + 4t_2$) as seen in the black curve in FIG.~\ref{fig:dos_vHs}. However as the nematic order parameter becomes nonzero, the saddle points at $(\pi,0)$ and $(0,\pi)$ lead to van Hove singularities in the bare DOS at different energies $\Tilde{\xi}_\vk^{vH} = \xi_\vk^{vH} - 2\Phi$ and $\Tilde{\xi}_\vk^{vH} = \xi_\vk^{vH} + 2\Phi$ respectively. This can be seen from the two singularities present in both the magenta and green curves in FIG.~\ref{fig:dos_vHs}. When the nematic order parameter reaches the critical value $\Phi_c = |-\frac{\mu}{2} + 2t_2|$, the van Hove singularities cross the Fermi level as indicated in the magenta curves in FIG.~\ref{fig:dos_vHs}. The van Hove singularities crossing the Fermi level \cite{Kreisel2021} has an  impact on the transport properties of the system when $T>T_c$ and will be discussed in Section \ref{sec:results}.
%~~~~~~~~~~~~~~~~~~~~~~~~~~~~~~~~~~~~~~~~~~~~~~~~~~~~~~~~~~~~~~~~~~~~~~~~

\begin{figure}
\begin{centering}
\includegraphics[width=8.5cm]{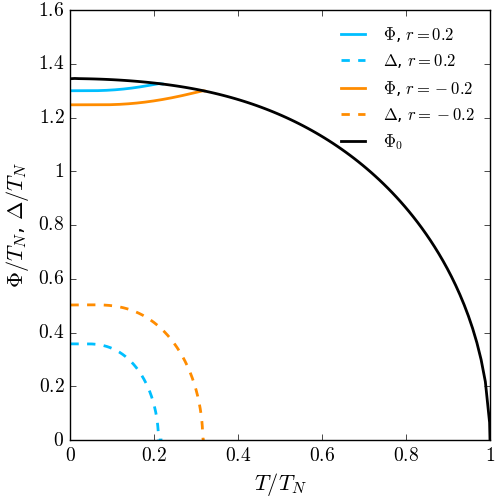}
\caption{Self-consistent solutions for $\Phi(T)$ and $\Delta(T)$ on the closed ($\mu = -4.8 T_N$) tight-binding Fermi surface when  $r = 0.2$ (cyan curves), $r = -0.2$ (orange curves), and  $T_c^0 = 0.4 T_N$. Also pictured is $\Phi_0(T)$, the nematic order in the absence of superconductivity (black curve). The band parameters are $t_1 = 6 T_N$ and $t_2 = -T_N$.}
\label{fig:deltaPhi1}
\end{centering}
\end{figure}

\subsection{Self-consistent equations for Nematicity and Superconductivity }
\label{sec:selfconsistency}
The self-consistent equations for  $\Phi$ and $\Delta$ are obtained by calculating the averages in (\ref{ham_nem}) and (\ref{ham_sc}), respectively \cite{chen2020} 
\begin{align}
\Phi &= V_{nem}\sum\limits_{\vk}\frac{f_\vk}{2} \Bigg[\frac{\xi_\vk + \Phi f_\vk}{E_\vk} \tanh{\frac{E_\vk}{2T}} -1 \Bigg]
\label{phi}
\\
\Delta &= -V_{sc}\Delta\sum\limits_{\vk}\frac{\mathcal{Y}^2_\vk}{2E_\vk} \tanh{\frac{E_{\vk}}{2T}} 
\label{delta}
\end{align}
The equation for  $\Phi$ in the pure nematic phase is obtained by setting $\Delta=0$ in equation (\ref{phi}) and leads to the following self-consistent equation
\begin{align}
\Phi &= V_{nem}\sum\limits_{\vk}\frac{f_\vk}{2} \Bigg[\tanh{\frac{\xi_\vk + \Phi f_\vk}{2T}} -1\Bigg]
\label{purephi}
\end{align}
The equation that determines the nematic transition temperature $T_N$ is obtained by setting $\Phi \rightarrow 0$ as $T\rightarrow T_N$ in equation (\ref{purephi}), yielding 
\begin{align}
1 = \frac{V_{nem}}{2}\sum\limits_{\vk}\frac{f^2_\vk}{2T_N} \Bigg[\sech^2{\frac{\xi_\vk }{2T_N}}\Bigg].
\label{TN_selfConsistency}
\end{align}
The superconducting transition temperature in the absence of nematicity ($T_c^0$) can be determined from equation (\ref{delta}) 
\begin{align}
    1 &= -V_{sc}\sum\limits_{\vk}\frac{\mathcal{Y}^2_\vk}{2\xi_\vk} \tanh{\frac{\xi_{\vk}}{2T_{c}^0}}.
    \label{deltaTc}
\end{align}
Note that in all the cases considered in this work, $T_c^0$ has been set to $0.4T_N$. However, the superconducting transition temperature ($T_c$) in the presence of the nematic order is different from $T_c^0$ as can be seen in FIG.~\ref{fig:deltaPhi1}.

\subsubsection{Numerical Solution of Self-Consistent equations}

Equations (\ref{phi}) and (\ref{delta}) can be solved self-consistently. For clarity, the parameters $V_{nem}$ and $V_{sc}$ are eliminated in favor of $T_N$ and $T_c^0$ using equations (\ref{TN_selfConsistency}) and (\ref{deltaTc}). Similarly, $\Phi_0(T)$ (the nematic order parameter in the absence of SC) can also be solved self-consistently from equation (\ref{purephi}) where $V_{nem}$ was again eliminated in favor of $T_N$ using equation (\ref{TN_selfConsistency}). The solutions $\Phi(T)$ and $\Delta(T)$ for $r = \pm 0.2$  are shown in FIG.~\ref{fig:deltaPhi1}. It can be seen that in the presence of SC, the nematic order parameter is slightly diminished from its value in the absence of SC (i.e. $\Phi(T) < \Phi_0(T)$ when $\Delta(T) \neq 0$). The SC transition temperature is also lower in the presence of nematicity ($T_c = 0.211 T_N$ for $ r = 0.2$, $T_c = 0.317 T_N$ for $r = -0.2$, and $T_c^0 = 0.4 T_N$), which is indicative of competing nematic and SC orders\cite{chen2020}. This was found to be the case for all parameter combinations studied in this work.

\subsection{Kinetic Method for Heat Conductivity}
\label{sec:kinetic}

We use the Boltzmann kinetic equation approach to calculate the thermal conductivity for the system with intertwined orders. This method was widely used to compute thermal conductivity, both in $s$-wave superconductors\cite{bardeen,gel}, as well as in unconventional superconductors\cite{mineev,arfi,arfi2,fritz}. The expression for the thermal conductivity for a superconductor in the Boltzmann kinetic approach is given by the expression \cite{mineev} 
\begin{align}
\kappa_{ij}&=-\frac{2}{T}\intk E^2_\vk {v}_{\vk,i}v_{\vk,j}\frac{\partial f^0_\vk}{\partial E}\tau_{\vk}
\label{tc}
\end{align}
where $f^{0}_\vk= \frac{1}{e^{E_\vk/T}+1}$ is the equilibrium Fermi-Dirac distribution function. The quasiparticle velocity is defined as 
\begin{align}
\mathbf{v}_\vk=\nabla_{\vk}E_\vk
\label{qpvel} 
\end{align} 
and the quasiparticle relaxation time is given by\cite{arfi} 
\begin{small}
\begin{align}
\tau^{-1}_\vk=N_{imp}\frac{2\pi}{\hbar}\intkprfbz |t_{\mathbf{k,k'}}|^2\delta (E_{\vk} - E_{\vk'})
\label{scat rate2}
\end{align} 
\end{small}  
where $t_{\vk,\vk'}$ is the amplitude for a single impurity to scatter a quasiparticle from the state with momentum $\vk $ and energy $E_{\vk}$ to the state with momentum $\vk'$ and energy $E_{\vk'}$ and $N_{imp}$ is the density of impurities. 

In order to determine the amplitude $t_{\vk,\vk'}$, we first write the impurity scattering Hamiltonian in the same Nambu basis as equation (\ref{ham full})
\begin{align}
\begin{split}
H_{imp}&=v_{imp}\sum_{\mathbf{k,k'},\sigma} c_{\vk' \sigma}^{\dagger}c_{\vk \sigma}\\
&=\frac{1}{2}\sum\limits_{\mathbf{k,k'}} \hat{\Psi}^{\dagger}_{\vk'\sigma}\hat{v}\hat{\Psi}_{\vk\sigma}    \\
\hat{v}&=v_{imp}\hat{\tau}_3
\label{himp}
\end{split} 
\end{align}
where $\hat{\tau}_3$ is the Pauli matrix in Nambu space and $v_{imp}$ is a non-magnetic isotropic impurity potential. The operators $c_{\vk,\sigma}^{\dagger}$ and $c_{\vk,\sigma}$, which create and destroy normal state particles, are related to the superconducting state quasiparticles $a_{\vk,\sigma}^{\dagger}$ and $a_{\vk,\sigma}$ by the Bogoliubov transformation
\begin{align}
    \hat{\Psi}_{\vk,\sigma} &= \hat{B}^{(\sigma)}_{\vk} \hat{A}_{\vk}\\
\hat{B}^{(\sigma)}_{\vk} &= 
\begin{pmatrix}
u_\vk & -v_\vk \\
v_\vk & u_\vk \\
\end{pmatrix}
\label{bogoliubov}
\end{align}
where $u_\vk = \frac{E_\vk + \xi_\vk}{\sqrt{(E_\vk+\xi_\vk)^2 + \Delta_\vk^2}}$, $v_\vk = \frac{\sigma\Delta_\vk}{\sqrt{(E_\vk+\xi_\vk)^2 + \Delta_\vk^2}}$, and $\hat{A}_{\vk}^{\dagger} = \begin{pmatrix} a_{\vk,\sigma'}^{\dagger}, a_{-\vk,-\sigma'}\end{pmatrix}$.
Upon performing the Bogoliubov transformation (\ref{bogoliubov}) on the Nambu vectors, we get 
\begin{equation}
H_{imp}=\frac{1}{2}\sum\limits_{\mathbf{k,k'}} \hat{A}^{\dagger}_{\vk}\hat{D}_{\vk,\vk'}\hat{A}_{\vk} 
\end{equation}
where the matrix $\hat{D}_{\vk,\vk'}$ is given by 
\begin{align}
\hat{D}_{\vk,\vk'} = (\hat{B}^{(\sigma)}_{\vk'})^{\dagger} \hat{v} \hat{B}^{(\sigma)}_{\vk}. 
\label{cmatrix}
\end{align}

Using this formalism, we can now determine some important terms. From the ordering of the $A^{\dagger}_\vk$ vector, the amplitude $t_{\vk,\vk'}$ in the Born approximation is given by 

\begin{equation}
    t_{\vk,\vk'} = (\hat{D}_{\vk,\vk'})_{11}
    \label{dmatrixelement}.
\end{equation}
To get the amplitude in the Unitary limit, we replace $\hat{v}$ in equation (\ref{cmatrix}) by the $T$-matrix for impurity scattering

\begin{equation}
    \hat{D}_{\vk,\vk'} = (\hat{B}^{(\sigma)}_{\vk'})^{\dagger} \hat{T} \hat{B}_{\vk}^{(\sigma)}.
    \label{dmatrix}
\end{equation}
The $T$-matrix can be obtained from\cite{arfi} the Lippmann-Schwinger equation
\begin{align}
    \hat{T} = \hat{v} + \hat{v}\sum_{\vk}\hat{G}_{\vk}(E)\hat{T}
    \label{lippmannSchwinger}
\end{align}
where $\hat{G}_{\vk}(E)$ is the single-particle Green's function for the superconductor in the absence of impurities, and is given by
\begin{align}
    \hat{G}_{\vk}(E) = \frac{1}{E^2-E_\vk^2} 
    \begin{pmatrix}
    E + \xi_\vk & \sigma\Delta_\vk \\
    \sigma\Delta_\vk & E - \xi_\vk
    \end{pmatrix}
    \label{greenFunction}
\end{align}

Using equation (\ref{greenFunction}) in equation (\ref{lippmannSchwinger}), we get
\begin{equation}
\hat{T}=\frac{v_{imp}\hat{\tau}_{3} + i v_{imp}^2\Tilde{N}_0 (g \hat{I}_{2\times 2} + h\hat{\tau}_1)}{1 + v_{imp}^2 \Tilde{N}_0^2(|g|^2-|h|^2)}
\label{tmatrix}    
\end{equation}
The functions $g(E_\vk)$ and $h(E_\vk)$ are given by 
\begin{align}
g(E_{\vk})&=-\frac{i}{\Tilde{N}_0} \sum_{\vk'} \frac{E_\vk}{E_\vk^2-E_{\vk'}^2} \\
h(E_{\vk})&=-\frac{i}{\Tilde{N}_0} \sum_{\vk'} \frac{\Delta_{\vk'}}{E_\vk^2-E_{\vk'}^2}
\label{gh}    
\end{align}
where $\Tilde{N}_0 \equiv N(\Tilde{\xi}_\vk=0)$ is the density of states on the FS deformed due to nematicity and therefore depends on $\Phi(T)$. As $T \rightarrow T_N$, $\Tilde{N}_0 = N_0$ where $N_0 \equiv N(\xi_\vk=0)$ which is the density of states on the original tight-binding FS. When $\Delta \rightarrow 0$, $g(E_\vk) = 1$ and $h(E_\vk) = 0$. The functions $g(E_\vk)$ and $h(E_\vk)$ are the normal and anomalous part of the quasiparticle self-energy respectively\cite{mineev}. The real part of the function $g(E_\vk)$ is proportional to the quasiparticle density of states and the imaginary part corresponds to dispersive corrections to the quasiparticle self-energy. 

The function $h(E_\vk)$ goes to zero for all superconducting states with the order parameters corresponding to non-identity representations of the crystal symmetry group (for example the $d_{x^2-y^2}$ and $d_{xy}$ pairing states \cite{mineev, arfi}). In our case $h(E_\vk) \neq 0$ due to the feedback from the symmetry broken nematic state on the SC order. The $T$-matrix in equation (\ref{tmatrix}) is directly  parameterized in terms of the strength of the impurity potential, $v_{imp}$, however it can also be  equivalently parameterized in terms of the normal state scattering phase shift $\delta_N$\cite{arfi}. In this paper we only consider two limiting cases: weak impurity potential ($v_{imp}\Tilde{N}_0 \ll 1 \Rightarrow  \delta_N \ll \pi/2$) which puts us in the limit where the Born approximation is valid, whereas a strong impurity potential ($v_{imp}\Tilde{N}_0 \gg 1 \Rightarrow \delta_N = \pi/2$) puts us in the Unitary limit. In the Born and Unitary limits, the $T$-matrix in equation (\ref{tmatrix}) reduces to
\begin{align}
\hat{T}_{Born}&=t_N^{Born}\hat{\tau}_{3} \label{tmatb}\\
\hat{T}_{Unitary}&=\frac{t_N^{Unitary}}{|g|^2-|h|^2}\big(g \hat{I}_{2\times2} + h \hat{\tau}_1\big)
\label{tmatu}
\end{align}
where $t_N^{Born} = v_{imp}$ and $t_N^{Unitary} = i/\Tilde{N}_0$.
Using equations (\ref{dmatrixelement}), (\ref{dmatrix}), (\ref{tmatb}), and (\ref{tmatu}) we can compute the amplitude $t_{\vk,\vk'}$ in the Born and Unitary limits respectively.
\begin{equation}
|t_{\vk,\vk'}|^2=\frac{|t_N^{Born}|^2}{2}\left(1+\frac{\xi_{\vk }\xi_{\vk' }-\Delta_{\vk }\Delta_{\vk' }}{E_\vk E_{\vk'}}\right)
\label{t2b}
\end{equation}
\begin{equation}
\begin{split}
|t_{\vk,\vk'}|^2=\frac{|t_N^{Unitary}|^2}{2}\bigg[a\bigg(1+\frac{\Delta_\vk \Delta_\vk'}{E_\vk E_\vk'}\bigg) \\
+ b\frac{\xi_\vk \xi_\vk'}{E_\vk E_\vk'} + 2c\bigg(\frac{\Delta_\vk}{E_\vk} + \frac{\Delta_\vk'}{E_\vk'}\bigg)\bigg]
\label{t2u}
\end{split}
\end{equation}
where $a,b$ and $c$ are defined as 
\begin{align}
a&=\frac{|g|^2 +|h|^2}{\big||g|^2 - |h|^2\big|^2}\\
b&=\frac{|g|^2 -|h|^2}{\big||g|^2 - |h|^2\big|^2}\\
c&=\frac{Re(g h^*)}{\big||g|^2 - |h|^2\big|^2}
\label{abc}    
\end{align}
Using equations (\ref{t2b}) and (\ref{t2u}) in equation (\ref{scat rate2}), the scattering rates in both the Born and Unitary limits respectively are found to be

\begin{equation}
    \tau_\vk^{-1} = \tau_{NF}^{-1} \bigg(Re(g(E_\vk)) - \frac{\Delta_\vk}{E_\vk}Re(h(E_\vk))\bigg)
    \label{scatRateBorn}
\end{equation}

\begin{equation}
\begin{split}
    \tau_\vk^{-1} = \tau_{NF}^{-1} \bigg\{a\bigg[Re(g(E_\vk)) + \frac{\Delta_\vk}{E_\vk}Re(h(E_\vk))\bigg] \\
    + 2c\bigg[\frac{\Delta_\vk}{E_\vk}Re(g(E_\vk))+Re(h(E_\vk)) \bigg]\bigg\}
    \label{scatRateUnitary}
\end{split}
\end{equation}
where $\tau_{NF}^{-1}$ is the scattering rate  on the nematically deformed FS  in the absence of the SC order and is defined as $\tau_N^{-1}(\Tilde{\xi}_\vk) = \frac{2\pi}{\hbar}N_{imp}|t_N|^2\Tilde{N}(\Tilde{\xi}_\vk)$, $\tau_{NF}^{-1}= \tau_N^{-1}(\Tilde{\xi}_\vk=0)$. In the Born and Unitary limits $t_N$ has been defined after equation (\ref{tmatu}) as $t_N^{Born}$ and $t_N^{Unitary}$. Note that when $\Delta = 0$, $a=b=1$ and $c=0$, and we find $\tau_\vk^{-1} = \tau_N^{-1}$ in both the Born and Unitary limits. Further, when $\Phi \rightarrow 0 \Rightarrow r \rightarrow 0 \Rightarrow \Delta_{\vk}=\Delta$ and $\tilde{N}_0=N_0$, $\tau_N= \tau_n$, $Re(g(E_\vk)) = N_{sc}(E_\vk)/N_0$, $h(E_\vk)=\frac{\Delta}{E_\vk} g(E_\vk)$, where $N_{sc}(E_\vk)$ is the quasiparticle DOS in the superconducting state. This reduces the quasiparticle scattering rate in equation (\ref{scatRateBorn}) to $\tau^{-1}_\vk = \tau_n^{-1} \frac{N_{sc}(E_\vk)}{N_0}\bigg(1 -  \frac{\Delta^2}{E^2_\vk}\bigg)$, which is the usual expression for an $s$-wave superconductor in the Born limit\cite{bardeen,mineev}. Again in the case when $\Phi \rightarrow 0 \Rightarrow r \rightarrow 0$ and $\Delta_\vk$ has $d_{x^2-y^2}$ symmetry, $h(E_\vk) = 0$ which implies $a = 1/|g|^2$ and $c = 0$. Therefore equation (\ref{scatRateBorn}) reduces to the well-known expression\cite{arfi,mineev}, $\tau_\vk^{-1} = \tau_n^{-1} \frac{N_{sc}(E_\vk)}{N_0}$, for the scattering rate of the $d_{x^2-y^2}$ pairing state in the Born limit. Furthermore equation (\ref{scatRateUnitary}) reduces to, $\tau^{-1}_\vk=\tau^{-1}_n\frac{N_{sc}(E_\vk)}{N_0}\frac{1}{|g(E_\vk)|^2}$, which is the scattering rate for the $d_{x^2-y^2}$ pairing state in the Unitary limit\cite{arfi,mineev}. Using equations (\ref{scatRateBorn}) and (\ref{scatRateUnitary}) we numerically compute the thermal conductivity tensor $\kappa_{ij}(T)$ from equation (\ref{tc}) in both the Born and Unitary limits. We also compute the conductivity in the purely nematic state $\kappa_{N}(T)$ by setting $\Delta = 0$ in equation (\ref{tc}), thus eliminating the unknowns $N_{imp}$ and $v_{imp}$ in favor of the nematic state relaxation time $\tau_N$. 

\section{Numerical Results and Discussion}
\label{sec:results}

\subsection{Pure nematic phase: $\Phi\neq 0$, $\Delta = 0$}

\begin{figure}[h!]
\begin{centering}
\includegraphics[width=8.5cm]{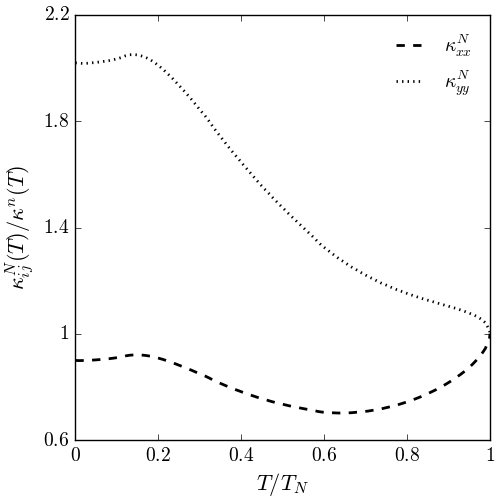}
\caption{Heat conductivity components ($\kappa_{ij}^N(T)$) of the nematically deformed closed FS with band parameters $t_1 = 6 T_N$, $t_2 = -T_N$, and $\mu = -4.8 T_N$ in the absence of SC order. $\kappa_{ij}^N(T)$ is normalized by the conductivity ($\kappa^n(T)$) of the normal state (original FS, $\Phi = 0$). The normal state conductivity is $T$-linear, $\kappa^n(T) = $ constant$\times T$.}
\label{fig:kappa_pureNem}
\end{centering}
\end{figure}

We begin our discussion by calculating the thermal conductivity of the pure nematic state for our tight-binding model with an initially closed Fermi surface. The components of the thermal conductivity tensor are normalized by the normal state ($\Phi = 0$ and $\Delta = 0$) conductivity ($\kappa^n(T)$). The results are shown in FIG.~\ref{fig:kappa_pureNem}, where we have treated the impurity scattering within the Born approximation. 

It can be seen that the $\kappa_{xx}^N$ and $\kappa_{yy}^N$ components of the thermal conductivity tensor are no longer equal, as is the case for the original ($\Phi = 0$) tight-binding Fermi surface (i.e $\kappa_{xx}^n = \kappa_{yy}^n = \kappa^n$ in the normal state). This is due to the fact that the nematic deformation has enhanced the quasiparticle velocities in the $y$-direction while diminishing the velocities in the $x$-direction (see FIG.~\ref{fig:nematicVelocity}). %Since velocities are orthogonal to the FS, this can be inferred from FIG~\ref{fig:FS}. The right and left corners of the original FS (black curve in FIG.~\ref{fig:FS}(a)) have velocities with components purely in the $x$-direction (i.e. $\mathbf{v}_{(k_F,0)} = (v_F,0)$). As the FS becomes deformed due to nematicity (green curve) these velocities become entirely in the $y$-direction (i.e. $\mathbf{v}_{(\pi,k_y^*)} = (0,\Tilde{v}_F)$, where $k_y^*$ lies on the green curve and $\Tilde{v}_F = \nabla \Tilde{\xi}_{\vk_F}$).
This results in $\kappa_{yy}^N$ always being greater than $\kappa_{xx}^N$. Despite these modifications to the quasiparticle velocities, the $\kappa_{xy}^N$ components still vanish due to the symmetry inherent in the velocities on the deformed FS.

While the effect of the nematic deformation on the Fermi velocities is an important characteristic, it cannot explain all the features of the thermal conductivity in FIG.~\ref{fig:kappa_pureNem}. If the nematic deformation only impacted the velocities as explained above, it would cause $\kappa_{yy}^N$ to increase by the same amount that $\kappa_{xx}^N$ decreases from $\kappa^n$, leading to a symmetric splitting in the $\kappa_{xx}^N$ and $\kappa_{yy}^N$ components. 

The asymmetric splitting in FIG.~\ref{fig:kappa_pureNem} is due to the fact that the particle lifetimes in the nematic state are different from the normal state. The particle lifetime in the nematic state is $\tau_N(\Tilde{\xi}_\vk) \propto 1/\Tilde{N}(\Tilde{\xi}_\vk)$ (defined below equation (\ref{scatRateUnitary})). In FIG.~\ref{fig:dos_vHs} it can be seen that as the van Hove singularities approach the Fermi level,  $\Tilde{N}(\Tilde{\xi}_\vk)$ near the Fermi level increases, which causes $\tau_N(\Tilde{\xi}_\vk)$ near the Fermi level to decrease. Thus, near $T_N$ (the van Hove singularities cross the Fermi level when $T = 0.97T_N$)  $\kappa_{xx}^N$ decreases much more quickly than $\kappa_{yy}^N$ increases. However after the van Hove singularity passes through the Fermi level, the DOS $\Tilde{N}(\Tilde{\xi}_\vk)$ near the Fermi level begins to decrease (see FIG.~\ref{fig:dos_vHs}), causing $\tau_N$ to increase. This results in long-lived, high velocity quasiparticles which conduct heat more efficiently, forcing $\kappa_{yy}^N$ to increase rather rapidly. 

Simultaneously, although the velocity of the quasiparticles  moving in the $x$-direction are reduced, the lifetimes are increased (which more than compensates for the velocity reduction), causing $\kappa_{xx}^N$ to also increase, but at a much slower rate than $\kappa_{yy}^N$. Finally, near $T = 0$, $\Phi(T)$ has reached saturation and remains at a constant value resulting in both the particle velocities and lifetimes becoming nearly constant at low T. This results in the usual metallic state with a conductivity that is linear in $T$. Thus, van Hove singularities crossing the Fermi level \cite{Kreisel2021} (due to FS deformations caused by nematicity) have a significant effect on the heat transport properties properties of the system when it is in the pure nematic phase.

\subsection{Pure Superconducting phase: $\Phi=0$, $\Delta \neq0$}
\label{puresc}
\begin{figure}[h!]
\begin{centering}
\includegraphics[width=8.5cm]{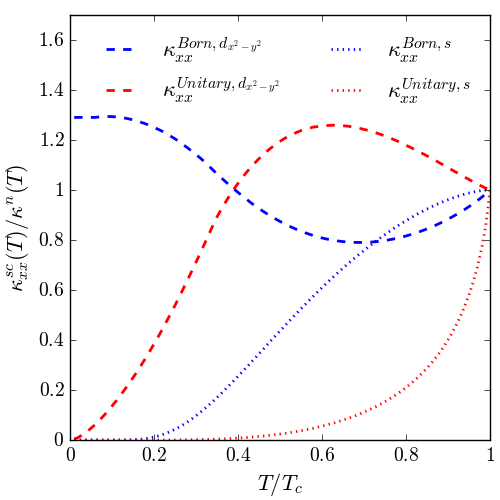}
\caption{Thermal conductivity components of the closed FS with band parameters $\mu = -48T_c$, $t_1=60T_c$, and $t_2=-10T_c$ in the pure $d_{x^2-y^2}$ and $s$-wave superconducting states in both the Born and Unitary limits.}
\label{fig:kappa_pureSC}
\end{centering}
\end{figure}
In FIG.~\ref{fig:kappa_pureSC} we have calculated the thermal conductivity of the pure SC states for our tight binding model. For the various pairing states, namely, $s$, $d_{x^2-y^2}$, the values of $\Delta(T)$ are obtained by self consistently solving the weak coupling gap equation. In the Born limit, we see the characteristic exponential fall in the thermal conductivity of the isotropic fully gapped $s$-wave superconductor \cite{bardeen}. 

The general behavior of $\kappa(T)/T$  in the Born limit, for the $d_{x^2-y^2}$ state also agrees with earlier calculations\cite{graf,arfi,choudhury2021}, where the low-$T$ regime is dominated by the nodal quasiparticles, producing a finite residual $\kappa/T$. The $d_{x^2-y^2}$ pairing has nodes on flat parts of the FS with large Fermi velocity and smaller DOS. By gapping the corners of the FS with large DOS, the scattering rate is significantly reduced, producing longer-lived high-velocity nodal quasiparticles that result in heat conductivity exceeding that of the normal state. The scattering rate in the pure $s$-wave state is given by the expression \cite{mineev} (see discussion below equation (\ref{scatRateUnitary})),  $\tau^{-1}_\vk = \tau_{N}^{-1}\frac{N_{sc}(E_\vk)}{N_{0}}\left( 1-\frac{\Delta^2}{E^2_\vk}\right)$. However, in the case of the pure $d_{x^2-y^2}$ state \cite{mineev}, $\tau^{-1}_\vk = \tau_{N}^{-1}\frac{N_{sc}(E_\vk)}{N_{0}}$. 

Comparing the coherence factors for various states, one can notice that near their transition temperatures the effective relaxation time for the $s$-wave state is greater than the $d_{x^2-y^2}$ state. This results in the observed different slopes near $T_c$ in FIG.~\ref{fig:kappa_pureSC} for the Born limit. 

In FIG.~\ref{fig:kappa_pureSC} we have also plotted thermal conductivity in the the unitary limit for both $s$ and $d_{x^2-y^2}$ pairing states. Again the general behaviour of $\kappa(T)/T$ in the unitary limit agrees with previously published results \cite{arfi,graf}.  The unitary limit result for the $d_{x^2-y^2}$ pairing state is in better agreement with experimental data for cuprates, than the Born approximation result. It has been found experimentally that at low temperatures $\kappa(T)$ has a power law like temperature dependence with an exponent greater than unity and that $\kappa(T) > \kappa_n(T)$ for intermediate temperatures \cite{yu1992, uher}.

%\begin{widetext}
\begin{figure*}[t]
\begin{centering}
\includegraphics[width=7in]{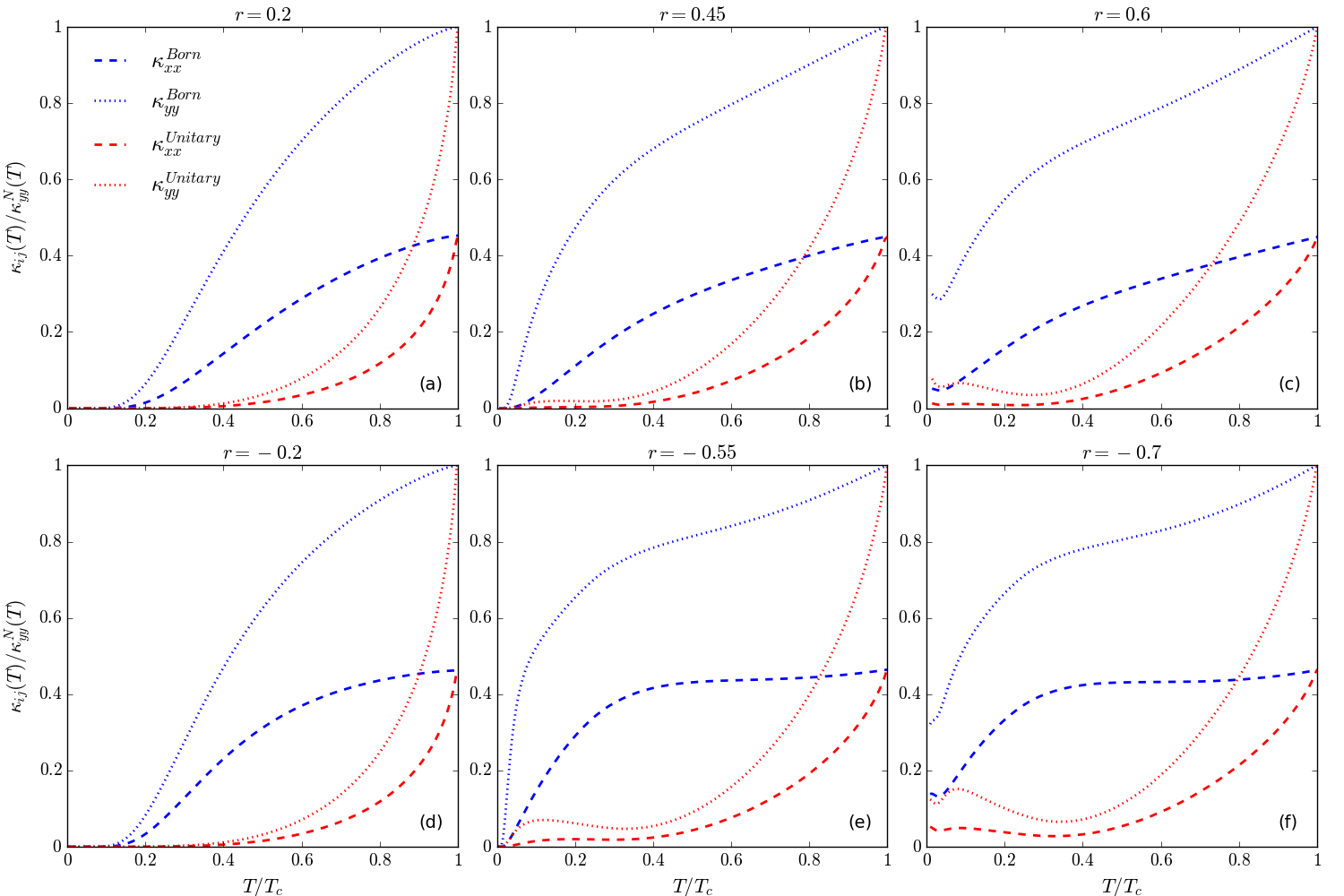}
\caption{Thermal conductivity components ($\kappa_{xx}$ and $\kappa_{yy}$) in the coexistence phase in both the Born (blue curves) and Unitary (red curves) limits normalized by the $\kappa_{yy}^N(T)$ component in the pure nematic phase when (a) $r = 0.2$ with $T_c = 0.211  T_N$, (b) $r = 0.45$ with $T_c = 0.158 T_N$, (c) $r = 0.6$ with $T_c = 0.140 T_N$, (d) $r = -0.2$ with $T_c = 0.317 T_N$, (e) $r = -0.55$ with $T_c = 0.332 T_N$, and (f) $r = -0.7$ with $T_c = 0.323 T_N$. The parameters used are  $\mu = -4.8 T_N$, $t_1 = 6 T_N$, $t_2 = - T_N$, and $\Phi = 1.34 T_N$. The critical $r$-values are $r_c^+ \approx 0.52866$ and $r_c^- \approx -0.61447$. Thus the FS (at $T=0$) in the coexistence phase corresponds to the green curve in FIG.~\ref{fig:FS}).}
\label{fig:kappa_all}
\end{centering}
\end{figure*}
%\end{widetext}

\subsection{Coexistence phase: $\Phi\neq 0$, $\Delta \neq 0$}

In this section and what follows, to study the effects of SC order emerging from a nematic background, we discuss the components of the thermal conductivity tensor and thermal transport in the coexistence phase, where the SC order and the nematic order are simultaneously nonzero. To illustrate important aspects of our results and emphasize the fact that $\kappa_{yy}$ is always greater than  $\kappa_{xx}$ when $\Phi \neq 0$, we have chosen to normalize $\kappa_{yy}(T)$ and $\kappa_{xx}(T)$ by the nematic state thermal conductivity component $\kappa_{yy}^N(T)$, as a result of which $\kappa_{yy}(T) = 1$ at $T = T_c$. Apart from the the distortion of the FS, the nematic order parameter $\Phi(T)$ has another important consequence which pertains to the coexistence phase. As previously discussed, the feedback from the symmetry broken nematic phase on the SC order leads to the mixing of the $s$-wave and $d$-wave channels. The degree of this mixing  is determined by the parameter $r \propto \Phi(T)$. Therefore, we categorize our study of thermal transport into three cases:
weak mixing ($|r| \ll |r_c^\pm|$), moderate mixing ($|r| \lesssim |r_c^\pm|$), and strong mixing ($|r| > |r_c^\pm|$), all displayed in FIG.~\ref{fig:kappa_all}. As before, we have computed the thermal conductivity using the Boltzmann transport equation method and treated the impurity scattering in both the Born and Unitary limits (as outlined in section \ref{sec:kinetic}).

There are certain common features in all the plots shown in FIG.~\ref{fig:kappa_all}. The conductivity components in the Born limit either fall to zero (FIG.~\ref{fig:kappa_all} (a), (b), (d), \& (e)) or to a residual value (FIG.~\ref{fig:kappa_all} (c) \& (f)). These changes occur significantly more slowly than the corresponding components in the Unitary limit due to the fact that, in the Unitary limit (which corresponds to strong scattering centers), the quasiparticles are significantly more short-lived than the Born limit (which corresponds to weak scattering centers). These longer-lived quasiparticles in the Born limit conduct heat more efficiently than those in the Unitary limit at lower temperatures.

Another common feature in FIG.~\ref{fig:kappa_all} is that when $r > 0$, $\kappa_{yy}^{Born}(T)$ falls roughly at the same rate as $\kappa_{xx}^{Born}(T)$ as $T$ decreases from $T_c$ relative to the conductivity in the pure nematic phase (see FIG.~\ref{fig:kappa_all}(a), (b), \& (c)). The slight difference in slope is because the Fermi velocity in the $x$ and $y$ directions are not equal.  For the case when $r < 0$, $\kappa_{xx}^{Born}(T)$ falls noticeably more slowly than $\kappa_{yy}^{Born}(T)$ for $T < T_c$ (see FIG.~\ref{fig:kappa_all}(d), (e), \& (f)) due to the correlation between the locations of the low-energy excitations in the BZ (when $r > 0$ as compared to when $r < 0$) and the Fermi velocities ($\Tilde{v}_{F,x}$ and $\Tilde{v}_{F,y}$) in the $x$- and $y$-directions along the nematically deformed FS. 

\begin{figure}[h!]
\begin{centering}
\includegraphics[width=8.5cm]{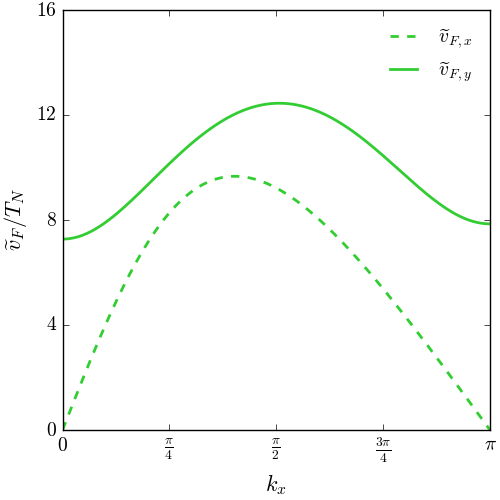}
\caption{Quasiparticle velocity in the pure nematic state plotted along the nematically deformed FS ($\Tilde{\mathbf{v}}_F = \nabla_\vk\Tilde{\xi}_{(k_x,k_y^*)}$) when $T = 0$ with band parameters $\mu = -4.8 T_N$, $t_1 = 6 T_N$, and $t_2 = - T_N$.}
\label{fig:nematicVelocity}
\end{centering}
\end{figure}

When $r > 0$, the low-energy excitations are located near $(\pm \pi,\pm k_y^*)$ whereas they are located near $(0,\pm k_y^*)$ for $r < 0$ (compare FIG.~\ref{fig:energyGap}(a) with FIG.~\ref{fig:energyGap}(b)). These low-energy excitations are primarily responsible for carrying the heat current in the coexistence phase. The quasiparticle velocities in the coexistence phase are $\mathbf{v}_\vk \approx \Tilde{\mathbf{v}}_F\frac{\Tilde{\xi}_{\vk}}{E_{\vk}}$, where $\Tilde{\mathbf{v}}_F$ is the Fermi velocity corresponding to the nematically deformed FS. In the regions around low-energy excitations for both $r > 0$ and $r < 0$, $\Tilde{v}_{F,y}$ are roughly equal and greater than $\Tilde{v}_{F,x}$ resulting in $\kappa_{yy}^{Born}$ being always greater than $\kappa_{xx}^{Born}$ with the slope of $\kappa^{Born}_{yy}$ being roughly equal for both $r > 0$ and $r < 0$. However, as seen in (see FIG.~\ref{fig:nematicVelocity}) the Fermi velocities in the $x$-direction are greater around the point $(0,\pm k_y^*)$ compared to $(\pm\pi,k_y^*)$, resulting in faster quasiparticles for $r>0$-values compared to when $r<0$. This results in $\kappa_{xx}^{Born}(T)$ to decrease more slowly near $T_c$ for $r < 0$ when compared to $r > 0$.

\subsubsection{weak mixing: $|r| \ll |r_c^\pm|$}
At low values of $r$ the SC gap is weakly anisotropic (for $r = 0.2$: $|\Delta_\vk|^{min}/|\Delta_\vk|^{max} = 0.46$ and, and for $r = -0.2$: $|\Delta_\vk|^{min}/|\Delta_\vk|^{max} = 0.48$ and ) and thus differs only slightly from the case of the uniformly gapped $s$-wave superconductor (see FIG.~ \ref{fig:compVcoop}(a) \& (d)). Therefore, in the case of weak mixing ($r=\pm 0.2$, see FIG.~\ref{fig:kappa_all} (a) \& (d)) the thermal conductivity profiles for both the components $\kappa_{xx}$ and $\kappa_{yy}$ are similar to the well known results for $s$-wave pairing \cite{bardeen, mineev} (compare FIG.~\ref{fig:kappa_all} (a) \& (d) to  FIG.~\ref{fig:kappa_pureSC}). Since $0.2 \ll r_c^+$ and $-0.2 \gg r_c^-$ for our chosen band parameters, no nodes exist in these cases and the FS is fully-gapped by the SC order. Therefore, there are only gapped excitaions in the coexistence phase, leading to an exponential reduction at low $T$ for both the Born and Unitary limits in FIG.~\ref{fig:kappa_all} (a) \& (d). 

\subsubsection{moderate mixing: $|r| \lesssim |r_c^\pm|$}

\begin{figure}
\begin{centering}
\includegraphics[width=7.8cm]{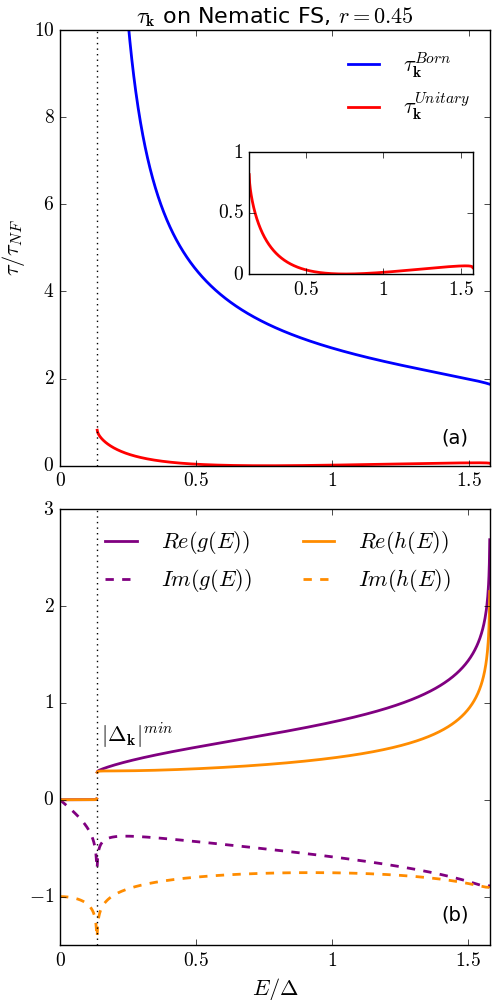}
\caption{(a) Quasiparticle lifetimes in the coexistence phase in the Born (blue curves) and Unitary (red curves) limits on the nematically deformed FS, normalized by the quasiparticle lifetimes on the FS ($\tau_{N F}$) in the pure nematic state. The inset is an expanded display of the lifetime in the Unitary limit. $E$ ranges from zero to $|\Delta_{\vk}|^{max}$, the maximum value of the gap amplitude on the FS. The black dotted line indicates the minimum value of the gap amplitude on the FS, $|\Delta_{\vk}|^{min}$. (b) The real and imaginary parts of $g(E)$ and $h(E)$ plotted over the same energies to illustrate their effects on the quasiparticle lifetimes.}
\label{fig:tau_moderate}
\end{centering}
\end{figure}
As the magnitude of $s$-wave and $d$-wave mixing is allowed to increase to $r = 0.45$ ($|\Delta_\vk|^{min}/|\Delta_\vk|^{max} = 0.08$) and $r = -0.55$ ($|\Delta_\vk|^{min}/|\Delta_\vk|^{max} = 0.05$), the SC gap develops deep minima on the FS (see FIG.~ \ref{fig:compVcoop}(b)\&(e) and FIG.~\ref{fig:energyGap}) with the resulting thermal conductivity profiles are displayed in FIG.~\ref{fig:kappa_all} (b) \& (e). The $d$-wave component in the SC order parameter becomes stronger as we transition from weak ($r=\pm 0.2 $) to moderate ($r=0.45, r=-0.55$) mixing, resulting in the effective relaxation time to decrease near $T_c$ in the Born limit, as explained previously in Section \ref{puresc}.  This change is reflected in the slopes near $T_c$ in FIG.~\ref{fig:kappa_all} (for the Born limit). Further, as the non-uniformity in the order parameter increases, the Fermi surface is no longer efficiently gapped by the SC order which results in the presence of excitations with lower energy than in the weak mixing case. Thus both thermal conductivity tensor components fall to $0$ at much lower temperatures compared to the weak mixing case. Unlike the $d$-wave state, $\kappa_{ij}$ components eventually fall to $0$ at low $T$ in the Born limit. This is a direct consequence of the fact that the system is still fully-gapped by the SC order (because $|r| \lesssim |r_c^\pm|$). 

In the Unitary limit, the lifetime $\tau_{\vk}$ at the FS begins to increase at low energies due to the stronger anisotropy in the SC gap and $\kappa_{ij}$ has a slight upturn before falling to zero at low $T$. Since the real part of $g(E)$ corresponds to the density of states in the coexistence phase, $Re(g(E))=0$ for $E<|\Delta_\vk|^{min}$, as there can be no excitations below the minimum value of the energy gap. Further, there is a coherence peak in the density of states at $E=|\Delta_\vk|^{max}$. As $E\rightarrow |\Delta_\vk|^{min}$ both the $Re(g(E))$ and $Re(h(E))$ decrease, whereas $Im(g(E))$ and $Im(h(E))$ increase, causing the parameters $a$ and $c$ to increase (see equation (\ref{abc})). This results in a reduction $\tau_\vk^{-1}$ (see equation (\ref{scatRateUnitary})) and a consequent increase in the quasiparticle lifetime in the unitary limit as $E\rightarrow |\Delta_\vk|^{min}$.

\subsubsection{strong mixing: $|r| > |r_c^\pm|$}

\begin{figure}[t]
\begin{centering}
\includegraphics[width=7.8cm]{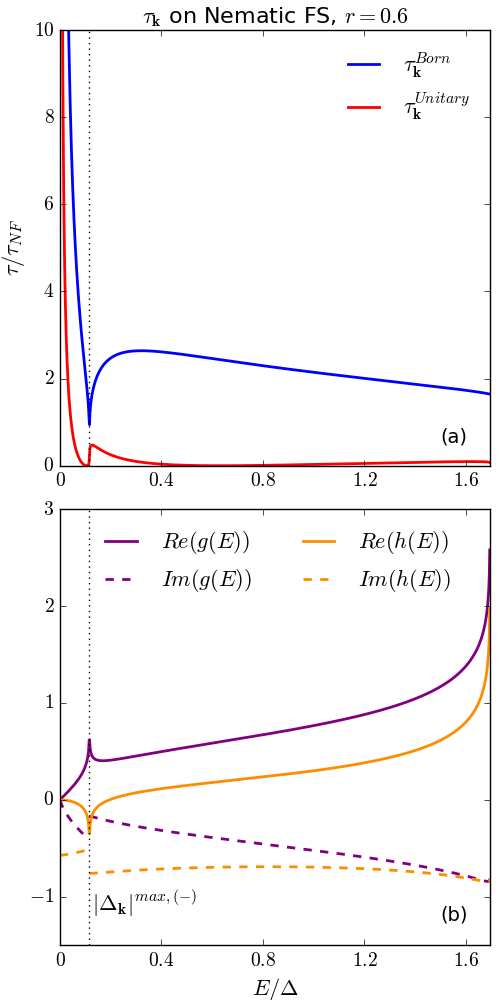}
\caption{(a) Quasiparticle lifetimes  in the coexistence phase in the Born (blue curves) and Unitary (red curves) limits on the nematically deformed FS, normalized by the quasiparticle lifetimes on the FS ($\tau_{N F}$) in the pure nematic state. $E$ ranges from zero to $|\Delta_{\vk}|^{max,(+)}$, the maximum value of the gap amplitude on the FS. The black line indicates the secondary, negative gap maximum on the FS, $|\Delta_{\vk}|^{max, (-)}$. (b) The real and imaginary parts of $g(E)$ and $h(E)$ plotted over the same energies to illustrate their effects on the quasiparticle lifetimes.}
\label{fig:tau_nodes}
\end{centering}
\end{figure}
Finally, as $|r| > |r_c^\pm|$, the SC gap collapses at the nodal points on the FS.  The non-uniformity of the gap results in smaller secondary SC gap maxima $|\Delta_{\vk}|^{max,(-)}$ on the FS (see FIG.~\ref{fig:energyGap} and \ref{fig:compVcoop}), corresponding to the negative sign of the SC gap function. %\blue{(for $r = 0.6$: $|\Delta_\vk|^{max,(+)} = 1.69\Delta$ and $|\Delta_\vk|^{max,(-)} = 0.12\Delta$, and for $r = -0.7$: $|\Delta|^{max,(+)} = 1.90\Delta$ and $|\Delta_\vk|^{max,(-)} = 0.11\Delta$)}. 
The corresponding thermal conductivity profiles are presented in FIG.~\ref{fig:kappa_all} (c) \& (f). In comparison with the moderate mixing case (FIG.~\ref{fig:kappa_all} (b) \& (e)), there is now a residual thermal conductivity at $T = 0$ (an obvious consequence of the existence of zero-energy excitations at the nodes).

Furthermore, in both the Born and Unitary limits, the residual $\kappa_{yy}$ values are roughly the same for $r < 0$ and $r >0 $, (see FIG.~\ref{fig:kappa_all}(c) \& (f)). This is again because the $y$-velocities of the quasiparticles are roughly the same at the locations of the nodes. However, in both the Born and Unitary limits, the residual values of $\kappa_{xx}$ when $r < 0$ are greater than when $r > 0$. When  $r < r_c^- < 0$ the nodes appear around $(0,\pm k_y^*)$ whereas when  $r > r_c^+ > 0$, the nodes appear around  $(\pm\pi,k_y^*)$ (see FIG.~\ref{fig:nodeLocations}(b)). As seen in (see FIG.~\ref{fig:nematicVelocity}) the Fermi velocities in the $x$-direction are greater around the point $(0,\pm k_y^*)$ compared to $(\pm\pi,k_y^*)$, resulting in faster nodal quasiparticles for negative $r$-values, which conduct heat more efficiently.

Unlike the pure nodal $d_{x^2-y^2}$ pairing state (see FIG.~\ref{fig:kappa_pureSC}), the components of $\kappa_{ij}$ in the Unitary limit no longer go to $0$ as $T \rightarrow 0$ because the quasiparticle lifetimes on the Fermi surface diverge at low energies (see FIG.~\ref{fig:tau_nodes}). In addition, the real part of $g(E)$ and $h(E)$ go to zero as $E\rightarrow 0$ causing the lifetime $\tau_\vk$ to diverge as $E\rightarrow 0$ for both the Born and Unitary limits. 

The singularity in the quasiparticle lifetime in the Born limit in FIG.~\ref{fig:tau_nodes}(a) occurs due to the coherence peak in the SC DOS (see FIG.~\ref{fig:tau_nodes}(b)) that appears at the energy corresponding to the smaller secondary SC gap maxima $|\Delta_{\vk}|^{max,(-)}$ on the FS (see FIG.~\ref{fig:compVcoop} and \ref{fig:energyGap}). Finally, at $E=|\Delta_{\vk}|^{max,(-)}$, $|g|\approx|h|$ which causes $\tau_\vk^{-1}$ to diverge and therefore the quasiparticle lifetime $\tau_\vk$ vanishes at that energy in the Unitary limit. 

In closing, we mention that for each of the cases studied above, the lifetimes for the anisotropic pairing states with positive values of $r$ are the roughly the same as those with negative values of $r$ because the spectrum of low-energy excitations of the quasiparticles are nearly the same for both (see FIG.~\ref{fig:energyGap}. At the location of these low-energy excitations (i.e near the gap minima or nodes), the magnitude of the Fermi velocities are roughly the same, implying that the local density of states at those locations are also nearly equal. As a result the quasiparticle lifetimes corresponding to SC pairing states with either positive or negative values of the anisotropy parameter $r$ do not differ much from one another. We have therefore not included the lifetime plots for negative values of $r$.

%~~~~~~~~~~~~~~~~~~~~~~~~~~~~~~~~~~~~~~~~~~~~~~~~~~~~~~~~~~~~~~~~~~~~~~~~

\section{Conclusion} 
\label{sec:conclusion}
We have considered a single band electronic system where a spin singlet superconducting order appears inside a nematic  phase. We treat both the orders at the mean-field level in a tight-binding square lattice with the nematic order being modelled as a $d$-wave Pomeranchuk type instability. The feedback from the symmetry-broken nematic phase on the SC order was accounted for through a mixing of the $s$-wave and $d$-wave channels which is controlled by a constant, phenomenological anisotropy parameter, $r$. Depending on the value of $r$, the gap function can display a deep minima (in the case of moderate mixing) or nodes (in the case of strong mixing). By determining the amplitudes of the SC and the nematic orders self-consistenly for all temperatures, the nature of the low energy exciations could be analysed showing that for $r > r_c^+(\Phi)$ or $r < r_c^-(\Phi)$, the spectrum has nodes which create a non-uniformity in the SC gap (a direct outcome of the interplay of the  FS distortion due to nematcity).  This non-uniformity results in inequivalent gap maxima at $|\Delta_{\vk}|^{max,(-)}$ and $|\Delta_{\vk}|^{max,(+)}$.

Temperature dependence of the electronic heat conductivity in the mixed SC and Nematic system was computed using the Boltzmann transport equation method, where the impurity scattering collision integral and quasiparticle lifetime were determined in both the Born and Unitary limits. We conclude that the nematic deformation of the FS results in $\kappa_{xx}(T)\neq \kappa_{yy}(T)$ and that there are significant differences in the thermal conductivity behavior in the coexistence phase that can distinguish between deep minima or nodes in the anisotropic SC gap structure. In the case of the SC gap having deep minima on the FS,  $\kappa \rightarrow 0$ as $T\rightarrow 0$ in both the Born and Unitary limits. In the case when the SC gap function has nodes, low-energy excitations lead to a finite residual $\kappa/T$ in the $T \to 0$ in both the Born and Unitary limits.

\section*{Acknowledgements}
This work supported by NSF Award No. 1809846 and through the NSF MonArk Quantum Foundry supported under Award No. 1906383.

\appendix

%%%%%%%%%%%%%%%%%%%%%%%%%%%%%%%%%%%%%%%%%%%%%%%%%%%%%%%%%%%%%%%%%%%
\bibliography{ref}

\begin{thebibliography}{63}
\expandafter\ifx\csname natexlab\endcsname\relax\def\natexlab#1{#1}\fi
\expandafter\ifx\csname bibnamefont\endcsname\relax
  \def\bibnamefont#1{#1}\fi
\expandafter\ifx\csname bibfnamefont\endcsname\relax
  \def\bibfnamefont#1{#1}\fi
\expandafter\ifx\csname citenamefont\endcsname\relax
  \def\citenamefont#1{#1}\fi
\expandafter\ifx\csname url\endcsname\relax
  \def\url#1{\texttt{#1}}\fi
\expandafter\ifx\csname urlprefix\endcsname\relax\def\urlprefix{URL }\fi
\providecommand{\bibinfo}[2]{#2}
\providecommand{\eprint}[2][]{\url{#2}}

\bibitem[{\citenamefont{Ziman}(1960)}]{ziman}
\bibinfo{author}{\bibfnamefont{J.}~\bibnamefont{Ziman}},
  \emph{\bibinfo{title}{Electrons and Phonons}} (\bibinfo{publisher}{Clarendon
  Press}, \bibinfo{address}{Oxford}, \bibinfo{year}{1960}).

\bibitem[{\citenamefont{Bardeen et~al.}(1959)\citenamefont{Bardeen, Rickayzen,
  and Tewordt}}]{bardeen}
\bibinfo{author}{\bibfnamefont{J.}~\bibnamefont{Bardeen}},
  \bibinfo{author}{\bibfnamefont{G.}~\bibnamefont{Rickayzen}},
  \bibnamefont{and} \bibinfo{author}{\bibfnamefont{L.}~\bibnamefont{Tewordt}},
  \bibinfo{journal}{Phys.Rev.} \textbf{\bibinfo{volume}{113}},
  \bibinfo{pages}{982} (\bibinfo{year}{1959}).

\bibitem[{\citenamefont{Pfleiderer}(2009)}]{hfrev}
\bibinfo{author}{\bibfnamefont{C.}~\bibnamefont{Pfleiderer}},
  \bibinfo{journal}{Rev. Mod. Phys.} \textbf{\bibinfo{volume}{81}},
  \bibinfo{pages}{1551} (\bibinfo{year}{2009}),
  \urlprefix\url{https://link.aps.org/doi/10.1103/RevModPhys.81.1551}.

\bibitem[{\citenamefont{Van~Harlingen}(1995)}]{harlin}
\bibinfo{author}{\bibfnamefont{D.~J.} \bibnamefont{Van~Harlingen}},
  \bibinfo{journal}{Rev. Mod. Phys.} \textbf{\bibinfo{volume}{67}},
  \bibinfo{pages}{515} (\bibinfo{year}{1995}),
  \urlprefix\url{https://link.aps.org/doi/10.1103/RevModPhys.67.515}.

\bibitem[{\citenamefont{Tsuei and Kirtley}(2000)}]{tsu}
\bibinfo{author}{\bibfnamefont{C.~C.} \bibnamefont{Tsuei}} \bibnamefont{and}
  \bibinfo{author}{\bibfnamefont{J.~R.} \bibnamefont{Kirtley}},
  \bibinfo{journal}{Rev. Mod. Phys.} \textbf{\bibinfo{volume}{72}},
  \bibinfo{pages}{969} (\bibinfo{year}{2000}),
  \urlprefix\url{https://link.aps.org/doi/10.1103/RevModPhys.72.969}.

\bibitem[{\citenamefont{Taillefer}(2010)}]{tail}
\bibinfo{author}{\bibfnamefont{L.}~\bibnamefont{Taillefer}},
  \bibinfo{journal}{Annual Review of Condensed Matter Physics}
  \textbf{\bibinfo{volume}{1}}, \bibinfo{pages}{51} (\bibinfo{year}{2010}),
  \urlprefix\url{https://doi.org/10.1146/annurev-conmatphys-070909-104117}.

\bibitem[{\citenamefont{Agterberg et~al.}(2020)\citenamefont{Agterberg, Davis,
  Edkins, Fradkin, Van~Harlingen, Kivelson, Lee, Radzihovsky, Tranquada, and
  Wang}}]{pair}
\bibinfo{author}{\bibfnamefont{D.~F.} \bibnamefont{Agterberg}},
  \bibinfo{author}{\bibfnamefont{J.~S.} \bibnamefont{Davis}},
  \bibinfo{author}{\bibfnamefont{S.~D.} \bibnamefont{Edkins}},
  \bibinfo{author}{\bibfnamefont{E.}~\bibnamefont{Fradkin}},
  \bibinfo{author}{\bibfnamefont{D.~J.} \bibnamefont{Van~Harlingen}},
  \bibinfo{author}{\bibfnamefont{S.~A.} \bibnamefont{Kivelson}},
  \bibinfo{author}{\bibfnamefont{P.~A.} \bibnamefont{Lee}},
  \bibinfo{author}{\bibfnamefont{L.}~\bibnamefont{Radzihovsky}},
  \bibinfo{author}{\bibfnamefont{J.~M.} \bibnamefont{Tranquada}},
  \bibnamefont{and} \bibinfo{author}{\bibfnamefont{Y.}~\bibnamefont{Wang}},
  \bibinfo{journal}{Annual Review of Condensed Matter Physics}
  \textbf{\bibinfo{volume}{11}}, \bibinfo{pages}{231} (\bibinfo{year}{2020}),
  \urlprefix\url{https://doi.org/10.1146/annurev-conmatphys-031119-050711}.

\bibitem[{\citenamefont{Wen and Li}(2011)}]{Wen}
\bibinfo{author}{\bibfnamefont{H.-H.} \bibnamefont{Wen}} \bibnamefont{and}
  \bibinfo{author}{\bibfnamefont{S.}~\bibnamefont{Li}},
  \bibinfo{journal}{Annual Review of Condensed Matter Physics}
  \textbf{\bibinfo{volume}{2}}, \bibinfo{pages}{121} (\bibinfo{year}{2011}),
  \urlprefix\url{https://doi.org/10.1146/annurev-conmatphys-062910-140518}.

\bibitem[{\citenamefont{Stewart}(2011)}]{stewart}
\bibinfo{author}{\bibfnamefont{G.~R.} \bibnamefont{Stewart}},
  \bibinfo{journal}{Rev. Mod. Phys.} \textbf{\bibinfo{volume}{83}},
  \bibinfo{pages}{1589} (\bibinfo{year}{2011}),
  \urlprefix\url{https://link.aps.org/doi/10.1103/RevModPhys.83.1589}.

\bibitem[{\citenamefont{Chubukov}(2012)}]{chubukov}
\bibinfo{author}{\bibfnamefont{A.}~\bibnamefont{Chubukov}},
  \bibinfo{journal}{Annual Review of Condensed Matter Physics}
  \textbf{\bibinfo{volume}{3}}, \bibinfo{pages}{57} (\bibinfo{year}{2012}),
  \urlprefix\url{https://doi.org/10.1146/annurev-conmatphys-020911-125055}.

\bibitem[{\citenamefont{Arfi and Pethick}(1988)}]{arfi}
\bibinfo{author}{\bibfnamefont{B.}~\bibnamefont{Arfi}} \bibnamefont{and}
  \bibinfo{author}{\bibfnamefont{C.~J.} \bibnamefont{Pethick}},
  \bibinfo{journal}{Phys. Rev. B} \textbf{\bibinfo{volume}{38}},
  \bibinfo{pages}{2312} (\bibinfo{year}{1988}),
  \urlprefix\url{https://link.aps.org/doi/10.1103/PhysRevB.38.2312}.

\bibitem[{\citenamefont{Hirschfeld et~al.}(1986)\citenamefont{Hirschfeld,
  Vollhardt, and Wölfle}}]{Hirsch}
\bibinfo{author}{\bibfnamefont{P.}~\bibnamefont{Hirschfeld}},
  \bibinfo{author}{\bibfnamefont{D.}~\bibnamefont{Vollhardt}},
  \bibnamefont{and} \bibinfo{author}{\bibfnamefont{P.}~\bibnamefont{Wölfle}},
  \bibinfo{journal}{Solid State Commun.} \textbf{\bibinfo{volume}{59}},
  \bibinfo{pages}{111} (\bibinfo{year}{1986}).

\bibitem[{\citenamefont{Scharnberg et~al.}(1986)\citenamefont{Scharnberg,
  Walker, Monien, Tewordt, and Klemm}}]{scharn}
\bibinfo{author}{\bibfnamefont{K.}~\bibnamefont{Scharnberg}},
  \bibinfo{author}{\bibfnamefont{D.}~\bibnamefont{Walker}},
  \bibinfo{author}{\bibfnamefont{H.}~\bibnamefont{Monien}},
  \bibinfo{author}{\bibfnamefont{L.}~\bibnamefont{Tewordt}}, \bibnamefont{and}
  \bibinfo{author}{\bibfnamefont{R.~A.} \bibnamefont{Klemm}},
  \bibinfo{journal}{Solid State Commun.} \textbf{\bibinfo{volume}{60}},
  \bibinfo{pages}{263} (\bibinfo{year}{1986}).

\bibitem[{\citenamefont{Monien et~al.}(1987)\citenamefont{Monien, Scharnberg,
  and Walker}}]{mon}
\bibinfo{author}{\bibfnamefont{H.}~\bibnamefont{Monien}},
  \bibinfo{author}{\bibfnamefont{K.}~\bibnamefont{Scharnberg}},
  \bibnamefont{and} \bibinfo{author}{\bibfnamefont{D.}~\bibnamefont{Walker}},
  \bibinfo{journal}{Solid State Commun.} \textbf{\bibinfo{volume}{63}},
  \bibinfo{pages}{535} (\bibinfo{year}{1987}).

\bibitem[{\citenamefont{Durst and Lee}(2000)}]{durst}
\bibinfo{author}{\bibfnamefont{A.~C.} \bibnamefont{Durst}} \bibnamefont{and}
  \bibinfo{author}{\bibfnamefont{P.~A.} \bibnamefont{Lee}},
  \bibinfo{journal}{Phys. Rev. B} \textbf{\bibinfo{volume}{62}},
  \bibinfo{pages}{1270} (\bibinfo{year}{2000}),
  \urlprefix\url{https://link.aps.org/doi/10.1103/PhysRevB.62.1270}.

\bibitem[{\citenamefont{Graf et~al.}(1996)\citenamefont{Graf, Yip, Sauls, and
  Rainer}}]{graf}
\bibinfo{author}{\bibfnamefont{M.~J.} \bibnamefont{Graf}},
  \bibinfo{author}{\bibfnamefont{S.-K.} \bibnamefont{Yip}},
  \bibinfo{author}{\bibfnamefont{J.~A.} \bibnamefont{Sauls}}, \bibnamefont{and}
  \bibinfo{author}{\bibfnamefont{D.}~\bibnamefont{Rainer}},
  \bibinfo{journal}{Phys.Rev.} \textbf{\bibinfo{volume}{53}},
  \bibinfo{pages}{15147} (\bibinfo{year}{1996}).

\bibitem[{\citenamefont{Matsuda et~al.}(2006)\citenamefont{Matsuda, Izawa, and
  Vekhter}}]{Matsuda}
\bibinfo{author}{\bibfnamefont{Y.}~\bibnamefont{Matsuda}},
  \bibinfo{author}{\bibfnamefont{K.}~\bibnamefont{Izawa}}, \bibnamefont{and}
  \bibinfo{author}{\bibfnamefont{I.}~\bibnamefont{Vekhter}},
  \bibinfo{journal}{Journal of Physics: Condensed Matter}
  \textbf{\bibinfo{volume}{18}}, \bibinfo{pages}{R705} (\bibinfo{year}{2006}),
  \urlprefix\url{https://doi.org/10.1088%2F0953-8984%2F18%2F44%2Fr01}.

\bibitem[{\citenamefont{Shakeripour et~al.}(2009)\citenamefont{Shakeripour,
  Petrovic, and Taillefer}}]{Shak}
\bibinfo{author}{\bibfnamefont{H.}~\bibnamefont{Shakeripour}},
  \bibinfo{author}{\bibfnamefont{C.}~\bibnamefont{Petrovic}}, \bibnamefont{and}
  \bibinfo{author}{\bibfnamefont{L.}~\bibnamefont{Taillefer}},
  \bibinfo{journal}{New Journal of Physics} \textbf{\bibinfo{volume}{11}},
  \bibinfo{pages}{055065} (\bibinfo{year}{2009}),
  \urlprefix\url{https://doi.org/10.1088%2F1367-2630%2F11%2F5%2F055065}.

\bibitem[{\citenamefont{Lake et~al.}(2002)\citenamefont{Lake, Rønnow,
  Christensen, Aeppli, McMorrow, Vorderwisch, Smeibidl, Sasagawa, Nohara, and
  Mason}}]{lake}
\bibinfo{author}{\bibfnamefont{B.}~\bibnamefont{Lake}},
  \bibinfo{author}{\bibfnamefont{H.}~\bibnamefont{Rønnow}},
  \bibinfo{author}{\bibfnamefont{N.}~\bibnamefont{Christensen}},
  \bibinfo{author}{\bibfnamefont{K.}~\bibnamefont{Aeppli},
  \bibfnamefont{G.and~Lefmann}}, \bibinfo{author}{\bibfnamefont{D.~F.}
  \bibnamefont{McMorrow}},
  \bibinfo{author}{\bibfnamefont{P.}~\bibnamefont{Vorderwisch}},
  \bibinfo{author}{\bibfnamefont{N.}~\bibnamefont{Smeibidl},
  \bibfnamefont{P.and~Mangkorntong}},
  \bibinfo{author}{\bibfnamefont{T.}~\bibnamefont{Sasagawa}},
  \bibinfo{author}{\bibfnamefont{H.}~\bibnamefont{Nohara},
  \bibfnamefont{M.and~Takagi}}, \bibnamefont{and}
  \bibinfo{author}{\bibfnamefont{T.~E.} \bibnamefont{Mason}},
  \bibinfo{journal}{Nature} \textbf{\bibinfo{volume}{415}},
  \bibinfo{pages}{219} (\bibinfo{year}{2002}).

\bibitem[{\citenamefont{Mathur et~al.}(1998)\citenamefont{Mathur, Grosche,
  Julian, Walker, Freye, Haselwimmer, and Lonzarich}}]{mathur}
\bibinfo{author}{\bibfnamefont{N.~D.} \bibnamefont{Mathur}},
  \bibinfo{author}{\bibfnamefont{F.~M.} \bibnamefont{Grosche}},
  \bibinfo{author}{\bibfnamefont{S.~R.} \bibnamefont{Julian}},
  \bibinfo{author}{\bibfnamefont{I.~R.} \bibnamefont{Walker}},
  \bibinfo{author}{\bibfnamefont{D.~M.} \bibnamefont{Freye}},
  \bibinfo{author}{\bibfnamefont{R.~K.~W.} \bibnamefont{Haselwimmer}},
  \bibnamefont{and} \bibinfo{author}{\bibfnamefont{G.~G.}
  \bibnamefont{Lonzarich}}, \bibinfo{journal}{Nature}
  \textbf{\bibinfo{volume}{394}}, \bibinfo{pages}{39} (\bibinfo{year}{1998}),
  ISSN \bibinfo{issn}{1476-4687},
  \urlprefix\url{https://doi.org/10.1038/27838}.

\bibitem[{\citenamefont{Badoux et~al.}(2016)\citenamefont{Badoux, Tabis,
  Lalibert{\'e}, Grissonnanche, Vignolle, Vignolles, B{\'e}ard, Bonn, Hardy,
  Liang et~al.}}]{Badoux2016}
\bibinfo{author}{\bibfnamefont{S.}~\bibnamefont{Badoux}},
  \bibinfo{author}{\bibfnamefont{W.}~\bibnamefont{Tabis}},
  \bibinfo{author}{\bibfnamefont{F.}~\bibnamefont{Lalibert{\'e}}},
  \bibinfo{author}{\bibfnamefont{G.}~\bibnamefont{Grissonnanche}},
  \bibinfo{author}{\bibfnamefont{B.}~\bibnamefont{Vignolle}},
  \bibinfo{author}{\bibfnamefont{D.}~\bibnamefont{Vignolles}},
  \bibinfo{author}{\bibfnamefont{J.}~\bibnamefont{B{\'e}ard}},
  \bibinfo{author}{\bibfnamefont{D.~A.} \bibnamefont{Bonn}},
  \bibinfo{author}{\bibfnamefont{W.~N.} \bibnamefont{Hardy}},
  \bibinfo{author}{\bibfnamefont{R.}~\bibnamefont{Liang}},
  \bibnamefont{et~al.}, \bibinfo{journal}{Nature}
  \textbf{\bibinfo{volume}{531}}, \bibinfo{pages}{210} (\bibinfo{year}{2016}),
  ISSN \bibinfo{issn}{1476-4687},
  \urlprefix\url{https://doi.org/10.1038/nature16983}.

\bibitem[{\citenamefont{Kim et~al.}(2016)\citenamefont{Kim, Lin, Weickert,
  Kenzelmann, Bauer, Ronning, Thompson, and Movshovich}}]{kim}
\bibinfo{author}{\bibfnamefont{D.~Y.} \bibnamefont{Kim}},
  \bibinfo{author}{\bibfnamefont{S.-Z.} \bibnamefont{Lin}},
  \bibinfo{author}{\bibfnamefont{F.}~\bibnamefont{Weickert}},
  \bibinfo{author}{\bibfnamefont{M.}~\bibnamefont{Kenzelmann}},
  \bibinfo{author}{\bibfnamefont{E.~D.} \bibnamefont{Bauer}},
  \bibinfo{author}{\bibfnamefont{F.}~\bibnamefont{Ronning}},
  \bibinfo{author}{\bibfnamefont{J.~D.} \bibnamefont{Thompson}},
  \bibnamefont{and}
  \bibinfo{author}{\bibfnamefont{R.}~\bibnamefont{Movshovich}},
  \bibinfo{journal}{Phys. Rev. X} \textbf{\bibinfo{volume}{6}},
  \bibinfo{pages}{041059} (\bibinfo{year}{2016}),
  \urlprefix\url{https://link.aps.org/doi/10.1103/PhysRevX.6.041059}.

\bibitem[{\citenamefont{Doiron-Leyraud
  et~al.}(2009)\citenamefont{Doiron-Leyraud, Auban-Senzier, Ren\'e~de Cotret,
  Bourbonnais, J\'erome, Bechgaard, and Taillefer}}]{ley}
\bibinfo{author}{\bibfnamefont{N.}~\bibnamefont{Doiron-Leyraud}},
  \bibinfo{author}{\bibfnamefont{P.}~\bibnamefont{Auban-Senzier}},
  \bibinfo{author}{\bibfnamefont{S.}~\bibnamefont{Ren\'e~de Cotret}},
  \bibinfo{author}{\bibfnamefont{C.}~\bibnamefont{Bourbonnais}},
  \bibinfo{author}{\bibfnamefont{D.}~\bibnamefont{J\'erome}},
  \bibinfo{author}{\bibfnamefont{K.}~\bibnamefont{Bechgaard}},
  \bibnamefont{and}
  \bibinfo{author}{\bibfnamefont{L.}~\bibnamefont{Taillefer}},
  \bibinfo{journal}{Phys. Rev. B} \textbf{\bibinfo{volume}{80}},
  \bibinfo{pages}{214531} (\bibinfo{year}{2009}),
  \urlprefix\url{https://link.aps.org/doi/10.1103/PhysRevB.80.214531}.

\bibitem[{\citenamefont{Chuang et~al.}(2010)\citenamefont{Chuang, Allan, Lee,
  Xie, Ni, Bud’ko, Boebinger, Canfield, and Davis}}]{chuang2010}
\bibinfo{author}{\bibfnamefont{T.-M.} \bibnamefont{Chuang}},
  \bibinfo{author}{\bibfnamefont{M.~P.} \bibnamefont{Allan}},
  \bibinfo{author}{\bibfnamefont{J.}~\bibnamefont{Lee}},
  \bibinfo{author}{\bibfnamefont{Y.}~\bibnamefont{Xie}},
  \bibinfo{author}{\bibfnamefont{N.}~\bibnamefont{Ni}},
  \bibinfo{author}{\bibfnamefont{S.~L.} \bibnamefont{Bud’ko}},
  \bibinfo{author}{\bibfnamefont{G.~S.} \bibnamefont{Boebinger}},
  \bibinfo{author}{\bibfnamefont{P.~C.} \bibnamefont{Canfield}},
  \bibnamefont{and} \bibinfo{author}{\bibfnamefont{J.~C.} \bibnamefont{Davis}},
  \bibinfo{journal}{Science} \textbf{\bibinfo{volume}{327}},
  \bibinfo{pages}{181} (\bibinfo{year}{2010}),
  \eprint{https://www.science.org/doi/pdf/10.1126/science.1181083},
  \urlprefix\url{https://www.science.org/doi/abs/10.1126/science.1181083}.

\bibitem[{\citenamefont{Böhmer and Meingast}(2016)}]{BOHMER201690}
\bibinfo{author}{\bibfnamefont{A.~E.} \bibnamefont{Böhmer}} \bibnamefont{and}
  \bibinfo{author}{\bibfnamefont{C.}~\bibnamefont{Meingast}},
  \bibinfo{journal}{Comptes Rendus Physique} \textbf{\bibinfo{volume}{17}},
  \bibinfo{pages}{90} (\bibinfo{year}{2016}), ISSN \bibinfo{issn}{1631-0705},
  \bibinfo{note}{iron-based superconductors / Supraconducteurs à base de fer},
  \urlprefix\url{https://www.sciencedirect.com/science/article/pii/S1631070515001279}.

\bibitem[{\citenamefont{Chu et~al.}(2012)\citenamefont{Chu, Kuo, Analytis, and
  Fisher}}]{chu2012}
\bibinfo{author}{\bibfnamefont{J.-H.} \bibnamefont{Chu}},
  \bibinfo{author}{\bibfnamefont{H.-H.} \bibnamefont{Kuo}},
  \bibinfo{author}{\bibfnamefont{J.~G.} \bibnamefont{Analytis}},
  \bibnamefont{and} \bibinfo{author}{\bibfnamefont{I.~R.}
  \bibnamefont{Fisher}}, \bibinfo{journal}{Science}
  \textbf{\bibinfo{volume}{337}}, \bibinfo{pages}{710} (\bibinfo{year}{2012}),
  \eprint{https://www.science.org/doi/pdf/10.1126/science.1221713},
  \urlprefix\url{https://www.science.org/doi/abs/10.1126/science.1221713}.

\bibitem[{\citenamefont{Nakata et~al.}(2021)\citenamefont{Nakata, Horio,
  Koshiishi, Hagiwara, Lin, Suzuki, Ideta, Tanaka, Song, Yoshida
  et~al.}}]{Nakata2021}
\bibinfo{author}{\bibfnamefont{S.}~\bibnamefont{Nakata}},
  \bibinfo{author}{\bibfnamefont{M.}~\bibnamefont{Horio}},
  \bibinfo{author}{\bibfnamefont{K.}~\bibnamefont{Koshiishi}},
  \bibinfo{author}{\bibfnamefont{K.}~\bibnamefont{Hagiwara}},
  \bibinfo{author}{\bibfnamefont{C.}~\bibnamefont{Lin}},
  \bibinfo{author}{\bibfnamefont{M.}~\bibnamefont{Suzuki}},
  \bibinfo{author}{\bibfnamefont{S.}~\bibnamefont{Ideta}},
  \bibinfo{author}{\bibfnamefont{K.}~\bibnamefont{Tanaka}},
  \bibinfo{author}{\bibfnamefont{D.}~\bibnamefont{Song}},
  \bibinfo{author}{\bibfnamefont{Y.}~\bibnamefont{Yoshida}},
  \bibnamefont{et~al.}, \bibinfo{journal}{npj Quantum Materials}
  \textbf{\bibinfo{volume}{6}}, \bibinfo{pages}{86} (\bibinfo{year}{2021}),
  ISSN \bibinfo{issn}{2397-4648},
  \urlprefix\url{https://doi.org/10.1038/s41535-021-00390-x}.

\bibitem[{\citenamefont{Ando et~al.}(2002)\citenamefont{Ando, Segawa, Komiya,
  and Lavrov}}]{ando2002}
\bibinfo{author}{\bibfnamefont{Y.}~\bibnamefont{Ando}},
  \bibinfo{author}{\bibfnamefont{K.}~\bibnamefont{Segawa}},
  \bibinfo{author}{\bibfnamefont{S.}~\bibnamefont{Komiya}}, \bibnamefont{and}
  \bibinfo{author}{\bibfnamefont{A.~N.} \bibnamefont{Lavrov}},
  \bibinfo{journal}{Phys. Rev. Lett.} \textbf{\bibinfo{volume}{88}},
  \bibinfo{pages}{137005} (\bibinfo{year}{2002}),
  \urlprefix\url{https://link.aps.org/doi/10.1103/PhysRevLett.88.137005}.

\bibitem[{\citenamefont{Hinkov et~al.}(2008)\citenamefont{Hinkov, Haug,
  Fauqué, Bourges, Sidis, Ivanov, Bernhard, Lin, and Keimer}}]{hinkov2008}
\bibinfo{author}{\bibfnamefont{V.}~\bibnamefont{Hinkov}},
  \bibinfo{author}{\bibfnamefont{D.}~\bibnamefont{Haug}},
  \bibinfo{author}{\bibfnamefont{B.}~\bibnamefont{Fauqué}},
  \bibinfo{author}{\bibfnamefont{P.}~\bibnamefont{Bourges}},
  \bibinfo{author}{\bibfnamefont{Y.}~\bibnamefont{Sidis}},
  \bibinfo{author}{\bibfnamefont{A.}~\bibnamefont{Ivanov}},
  \bibinfo{author}{\bibfnamefont{C.}~\bibnamefont{Bernhard}},
  \bibinfo{author}{\bibfnamefont{C.~T.} \bibnamefont{Lin}}, \bibnamefont{and}
  \bibinfo{author}{\bibfnamefont{B.}~\bibnamefont{Keimer}},
  \bibinfo{journal}{Science} \textbf{\bibinfo{volume}{319}},
  \bibinfo{pages}{597} (\bibinfo{year}{2008}),
  \eprint{https://www.science.org/doi/pdf/10.1126/science.1152309},
  \urlprefix\url{https://www.science.org/doi/abs/10.1126/science.1152309}.

\bibitem[{\citenamefont{Sato et~al.}(2017)\citenamefont{Sato, Kasahara,
  Murayama, Kasahara, Moon, Nishizaki, Loew, Porras, Keimer, Shibauchi
  et~al.}}]{Sato2017}
\bibinfo{author}{\bibfnamefont{Y.}~\bibnamefont{Sato}},
  \bibinfo{author}{\bibfnamefont{S.}~\bibnamefont{Kasahara}},
  \bibinfo{author}{\bibfnamefont{H.}~\bibnamefont{Murayama}},
  \bibinfo{author}{\bibfnamefont{Y.}~\bibnamefont{Kasahara}},
  \bibinfo{author}{\bibfnamefont{E.-G.} \bibnamefont{Moon}},
  \bibinfo{author}{\bibfnamefont{T.}~\bibnamefont{Nishizaki}},
  \bibinfo{author}{\bibfnamefont{T.}~\bibnamefont{Loew}},
  \bibinfo{author}{\bibfnamefont{J.}~\bibnamefont{Porras}},
  \bibinfo{author}{\bibfnamefont{B.}~\bibnamefont{Keimer}},
  \bibinfo{author}{\bibfnamefont{T.}~\bibnamefont{Shibauchi}},
  \bibnamefont{et~al.}, \bibinfo{journal}{Nature Physics}
  \textbf{\bibinfo{volume}{13}}, \bibinfo{pages}{1074} (\bibinfo{year}{2017}),
  ISSN \bibinfo{issn}{1745-2481},
  \urlprefix\url{https://doi.org/10.1038/nphys4205}.

\bibitem[{\citenamefont{Cyr-Choini\`ere
  et~al.}(2015)\citenamefont{Cyr-Choini\`ere, Grissonnanche, Badoux, Day, Bonn,
  Hardy, Liang, Doiron-Leyraud, and Taillefer}}]{cyr2015}
\bibinfo{author}{\bibfnamefont{O.}~\bibnamefont{Cyr-Choini\`ere}},
  \bibinfo{author}{\bibfnamefont{G.}~\bibnamefont{Grissonnanche}},
  \bibinfo{author}{\bibfnamefont{S.}~\bibnamefont{Badoux}},
  \bibinfo{author}{\bibfnamefont{J.}~\bibnamefont{Day}},
  \bibinfo{author}{\bibfnamefont{D.~A.} \bibnamefont{Bonn}},
  \bibinfo{author}{\bibfnamefont{W.~N.} \bibnamefont{Hardy}},
  \bibinfo{author}{\bibfnamefont{R.}~\bibnamefont{Liang}},
  \bibinfo{author}{\bibfnamefont{N.}~\bibnamefont{Doiron-Leyraud}},
  \bibnamefont{and}
  \bibinfo{author}{\bibfnamefont{L.}~\bibnamefont{Taillefer}},
  \bibinfo{journal}{Phys. Rev. B} \textbf{\bibinfo{volume}{92}},
  \bibinfo{pages}{224502} (\bibinfo{year}{2015}),
  \urlprefix\url{https://link.aps.org/doi/10.1103/PhysRevB.92.224502}.

\bibitem[{\citenamefont{Wu et~al.}(2017)\citenamefont{Wu, Bollinger, He, and
  Bo{\v{z}}ovi{\'{c}}}}]{Wu2017}
\bibinfo{author}{\bibfnamefont{J.}~\bibnamefont{Wu}},
  \bibinfo{author}{\bibfnamefont{A.~T.} \bibnamefont{Bollinger}},
  \bibinfo{author}{\bibfnamefont{X.}~\bibnamefont{He}}, \bibnamefont{and}
  \bibinfo{author}{\bibfnamefont{I.}~\bibnamefont{Bo{\v{z}}ovi{\'{c}}}},
  \bibinfo{journal}{Nature} \textbf{\bibinfo{volume}{547}},
  \bibinfo{pages}{432} (\bibinfo{year}{2017}), ISSN \bibinfo{issn}{1476-4687},
  \urlprefix\url{https://doi.org/10.1038/nature23290}.

\bibitem[{\citenamefont{Fernandes et~al.}(2014)\citenamefont{Fernandes,
  Chubukov, and Schmalian}}]{Fernandes2014}
\bibinfo{author}{\bibfnamefont{R.~M.} \bibnamefont{Fernandes}},
  \bibinfo{author}{\bibfnamefont{A.~V.} \bibnamefont{Chubukov}},
  \bibnamefont{and}
  \bibinfo{author}{\bibfnamefont{J.}~\bibnamefont{Schmalian}},
  \bibinfo{journal}{Nature Physics} \textbf{\bibinfo{volume}{10}},
  \bibinfo{pages}{97} (\bibinfo{year}{2014}), ISSN \bibinfo{issn}{1745-2481},
  \urlprefix\url{https://doi.org/10.1038/nphys2877}.

\bibitem[{\citenamefont{Hu and Xu}(2012)}]{HU2012215}
\bibinfo{author}{\bibfnamefont{J.}~\bibnamefont{Hu}} \bibnamefont{and}
  \bibinfo{author}{\bibfnamefont{C.}~\bibnamefont{Xu}},
  \bibinfo{journal}{Physica C: Superconductivity}
  \textbf{\bibinfo{volume}{481}}, \bibinfo{pages}{215} (\bibinfo{year}{2012}),
  ISSN \bibinfo{issn}{0921-4534}, \bibinfo{note}{stripes and Electronic Liquid
  Crystals in Strongly Correlated Materials},
  \urlprefix\url{https://www.sciencedirect.com/science/article/pii/S0921453412002341}.

\bibitem[{\citenamefont{Fernandes et~al.}(2013)\citenamefont{Fernandes,
  B\"ohmer, Meingast, and Schmalian}}]{Fernandes2013-2}
\bibinfo{author}{\bibfnamefont{R.~M.} \bibnamefont{Fernandes}},
  \bibinfo{author}{\bibfnamefont{A.~E.} \bibnamefont{B\"ohmer}},
  \bibinfo{author}{\bibfnamefont{C.}~\bibnamefont{Meingast}}, \bibnamefont{and}
  \bibinfo{author}{\bibfnamefont{J.}~\bibnamefont{Schmalian}},
  \bibinfo{journal}{Phys. Rev. Lett.} \textbf{\bibinfo{volume}{111}},
  \bibinfo{pages}{137001} (\bibinfo{year}{2013}),
  \urlprefix\url{https://link.aps.org/doi/10.1103/PhysRevLett.111.137001}.

\bibitem[{\citenamefont{B\"ohmer et~al.}(2013)\citenamefont{B\"ohmer, Hardy,
  Eilers, Ernst, Adelmann, Schweiss, Wolf, and Meingast}}]{Bohmer2013}
\bibinfo{author}{\bibfnamefont{A.~E.} \bibnamefont{B\"ohmer}},
  \bibinfo{author}{\bibfnamefont{F.}~\bibnamefont{Hardy}},
  \bibinfo{author}{\bibfnamefont{F.}~\bibnamefont{Eilers}},
  \bibinfo{author}{\bibfnamefont{D.}~\bibnamefont{Ernst}},
  \bibinfo{author}{\bibfnamefont{P.}~\bibnamefont{Adelmann}},
  \bibinfo{author}{\bibfnamefont{P.}~\bibnamefont{Schweiss}},
  \bibinfo{author}{\bibfnamefont{T.}~\bibnamefont{Wolf}}, \bibnamefont{and}
  \bibinfo{author}{\bibfnamefont{C.}~\bibnamefont{Meingast}},
  \bibinfo{journal}{Phys. Rev. B} \textbf{\bibinfo{volume}{87}},
  \bibinfo{pages}{180505} (\bibinfo{year}{2013}),
  \urlprefix\url{https://link.aps.org/doi/10.1103/PhysRevB.87.180505}.

\bibitem[{\citenamefont{Yamakawa et~al.}(2016)\citenamefont{Yamakawa, Onari,
  and Kontani}}]{yamakawa2016}
\bibinfo{author}{\bibfnamefont{Y.}~\bibnamefont{Yamakawa}},
  \bibinfo{author}{\bibfnamefont{S.}~\bibnamefont{Onari}}, \bibnamefont{and}
  \bibinfo{author}{\bibfnamefont{H.}~\bibnamefont{Kontani}},
  \bibinfo{journal}{Phys. Rev. X} \textbf{\bibinfo{volume}{6}},
  \bibinfo{pages}{021032} (\bibinfo{year}{2016}),
  \urlprefix\url{https://link.aps.org/doi/10.1103/PhysRevX.6.021032}.

\bibitem[{\citenamefont{Fanfarillo et~al.}(2018)\citenamefont{Fanfarillo,
  Benfatto, and Valenzuela}}]{fanfarillo2018}
\bibinfo{author}{\bibfnamefont{L.}~\bibnamefont{Fanfarillo}},
  \bibinfo{author}{\bibfnamefont{L.}~\bibnamefont{Benfatto}}, \bibnamefont{and}
  \bibinfo{author}{\bibfnamefont{B.}~\bibnamefont{Valenzuela}},
  \bibinfo{journal}{Phys. Rev. B} \textbf{\bibinfo{volume}{97}},
  \bibinfo{pages}{121109} (\bibinfo{year}{2018}),
  \urlprefix\url{https://link.aps.org/doi/10.1103/PhysRevB.97.121109}.

\bibitem[{\citenamefont{Kivelson et~al.}(2003)\citenamefont{Kivelson, Bindloss,
  Fradkin, Oganesyan, Tranquada, Kapitulnik, and Howald}}]{kivelson2003}
\bibinfo{author}{\bibfnamefont{S.~A.} \bibnamefont{Kivelson}},
  \bibinfo{author}{\bibfnamefont{I.~P.} \bibnamefont{Bindloss}},
  \bibinfo{author}{\bibfnamefont{E.}~\bibnamefont{Fradkin}},
  \bibinfo{author}{\bibfnamefont{V.}~\bibnamefont{Oganesyan}},
  \bibinfo{author}{\bibfnamefont{J.~M.} \bibnamefont{Tranquada}},
  \bibinfo{author}{\bibfnamefont{A.}~\bibnamefont{Kapitulnik}},
  \bibnamefont{and} \bibinfo{author}{\bibfnamefont{C.}~\bibnamefont{Howald}},
  \bibinfo{journal}{Rev. Mod. Phys.} \textbf{\bibinfo{volume}{75}},
  \bibinfo{pages}{1201} (\bibinfo{year}{2003}),
  \urlprefix\url{https://link.aps.org/doi/10.1103/RevModPhys.75.1201}.

\bibitem[{\citenamefont{Fradkin et~al.}(2010)\citenamefont{Fradkin, Kivelson,
  Lawler, Eisenstein, and Mackenzie}}]{fradkin2010}
\bibinfo{author}{\bibfnamefont{E.}~\bibnamefont{Fradkin}},
  \bibinfo{author}{\bibfnamefont{S.~A.} \bibnamefont{Kivelson}},
  \bibinfo{author}{\bibfnamefont{M.~J.} \bibnamefont{Lawler}},
  \bibinfo{author}{\bibfnamefont{J.~P.} \bibnamefont{Eisenstein}},
  \bibnamefont{and} \bibinfo{author}{\bibfnamefont{A.~P.}
  \bibnamefont{Mackenzie}}, \bibinfo{journal}{Annual Review of Condensed Matter
  Physics} \textbf{\bibinfo{volume}{1}}, \bibinfo{pages}{153}
  (\bibinfo{year}{2010}),
  \eprint{https://doi.org/10.1146/annurev-conmatphys-070909-103925},
  \urlprefix\url{https://doi.org/10.1146/annurev-conmatphys-070909-103925}.

\bibitem[{\citenamefont{Yamase and Metzner}(2006)}]{yamase2006}
\bibinfo{author}{\bibfnamefont{H.}~\bibnamefont{Yamase}} \bibnamefont{and}
  \bibinfo{author}{\bibfnamefont{W.}~\bibnamefont{Metzner}},
  \bibinfo{journal}{Phys. Rev. B} \textbf{\bibinfo{volume}{73}},
  \bibinfo{pages}{214517} (\bibinfo{year}{2006}),
  \urlprefix\url{https://link.aps.org/doi/10.1103/PhysRevB.73.214517}.

\bibitem[{\citenamefont{Oganesyan et~al.}(2001)\citenamefont{Oganesyan,
  Kivelson, and Fradkin}}]{oganesyan2001}
\bibinfo{author}{\bibfnamefont{V.}~\bibnamefont{Oganesyan}},
  \bibinfo{author}{\bibfnamefont{S.~A.} \bibnamefont{Kivelson}},
  \bibnamefont{and} \bibinfo{author}{\bibfnamefont{E.}~\bibnamefont{Fradkin}},
  \bibinfo{journal}{Phys. Rev. B} \textbf{\bibinfo{volume}{64}},
  \bibinfo{pages}{195109} (\bibinfo{year}{2001}),
  \urlprefix\url{https://link.aps.org/doi/10.1103/PhysRevB.64.195109}.

\bibitem[{\citenamefont{Kao and Kee}(2005)}]{kao2005}
\bibinfo{author}{\bibfnamefont{Y.-J.} \bibnamefont{Kao}} \bibnamefont{and}
  \bibinfo{author}{\bibfnamefont{H.-Y.} \bibnamefont{Kee}},
  \bibinfo{journal}{Phys. Rev. B} \textbf{\bibinfo{volume}{72}},
  \bibinfo{pages}{024502} (\bibinfo{year}{2005}),
  \urlprefix\url{https://link.aps.org/doi/10.1103/PhysRevB.72.024502}.

\bibitem[{\citenamefont{Halboth and Metzner}(2000)}]{halboth2000}
\bibinfo{author}{\bibfnamefont{C.~J.} \bibnamefont{Halboth}} \bibnamefont{and}
  \bibinfo{author}{\bibfnamefont{W.}~\bibnamefont{Metzner}},
  \bibinfo{journal}{Phys. Rev. Lett.} \textbf{\bibinfo{volume}{85}},
  \bibinfo{pages}{5162} (\bibinfo{year}{2000}),
  \urlprefix\url{https://link.aps.org/doi/10.1103/PhysRevLett.85.5162}.

\bibitem[{\citenamefont{Fernandes and Millis}(2013)}]{Fernandes2013}
\bibinfo{author}{\bibfnamefont{R.~M.} \bibnamefont{Fernandes}}
  \bibnamefont{and} \bibinfo{author}{\bibfnamefont{A.~J.}
  \bibnamefont{Millis}}, \bibinfo{journal}{Phys. Rev. Lett.}
  \textbf{\bibinfo{volume}{111}}, \bibinfo{pages}{127001}
  (\bibinfo{year}{2013}),
  \urlprefix\url{https://link.aps.org/doi/10.1103/PhysRevLett.111.127001}.

\bibitem[{\citenamefont{Chen et~al.}(2020)\citenamefont{Chen, Maiti, Fernandes,
  and Hirschfeld}}]{chen2020}
\bibinfo{author}{\bibfnamefont{X.}~\bibnamefont{Chen}},
  \bibinfo{author}{\bibfnamefont{S.}~\bibnamefont{Maiti}},
  \bibinfo{author}{\bibfnamefont{R.~M.} \bibnamefont{Fernandes}},
  \bibnamefont{and} \bibinfo{author}{\bibfnamefont{P.~J.}
  \bibnamefont{Hirschfeld}}, \bibinfo{journal}{Phys. Rev. B}
  \textbf{\bibinfo{volume}{102}}, \bibinfo{pages}{184512}
  (\bibinfo{year}{2020}),
  \urlprefix\url{https://link.aps.org/doi/10.1103/PhysRevB.102.184512}.

\bibitem[{\citenamefont{Fritz and Sachdev}(2009)}]{fritz}
\bibinfo{author}{\bibfnamefont{L.}~\bibnamefont{Fritz}} \bibnamefont{and}
  \bibinfo{author}{\bibfnamefont{S.}~\bibnamefont{Sachdev}},
  \bibinfo{journal}{Phys. Rev. B} \textbf{\bibinfo{volume}{80}},
  \bibinfo{pages}{144503} (\bibinfo{year}{2009}),
  \urlprefix\url{https://link.aps.org/doi/10.1103/PhysRevB.80.144503}.

\bibitem[{\citenamefont{Vorontsov and Vekhter}(2010)}]{vorontsov2010}
\bibinfo{author}{\bibfnamefont{A.~B.} \bibnamefont{Vorontsov}}
  \bibnamefont{and} \bibinfo{author}{\bibfnamefont{I.}~\bibnamefont{Vekhter}},
  \bibinfo{journal}{Phys. Rev. Lett.} \textbf{\bibinfo{volume}{105}},
  \bibinfo{pages}{187004} (\bibinfo{year}{2010}),
  \urlprefix\url{https://link.aps.org/doi/10.1103/PhysRevLett.105.187004}.

\bibitem[{\citenamefont{Chubukov and Eremin}(2010)}]{chubukov2010}
\bibinfo{author}{\bibfnamefont{A.~V.} \bibnamefont{Chubukov}} \bibnamefont{and}
  \bibinfo{author}{\bibfnamefont{I.}~\bibnamefont{Eremin}},
  \bibinfo{journal}{Phys. Rev. B} \textbf{\bibinfo{volume}{82}},
  \bibinfo{pages}{060504} (\bibinfo{year}{2010}),
  \urlprefix\url{https://link.aps.org/doi/10.1103/PhysRevB.82.060504}.

\bibitem[{\citenamefont{Hardy et~al.}(2019)\citenamefont{Hardy, He, Wang, Wolf,
  Schweiss, Merz, Barth, Adelmann, Eder, Haghighirad et~al.}}]{hardy2019}
\bibinfo{author}{\bibfnamefont{F.}~\bibnamefont{Hardy}},
  \bibinfo{author}{\bibfnamefont{M.}~\bibnamefont{He}},
  \bibinfo{author}{\bibfnamefont{L.}~\bibnamefont{Wang}},
  \bibinfo{author}{\bibfnamefont{T.}~\bibnamefont{Wolf}},
  \bibinfo{author}{\bibfnamefont{P.}~\bibnamefont{Schweiss}},
  \bibinfo{author}{\bibfnamefont{M.}~\bibnamefont{Merz}},
  \bibinfo{author}{\bibfnamefont{M.}~\bibnamefont{Barth}},
  \bibinfo{author}{\bibfnamefont{P.}~\bibnamefont{Adelmann}},
  \bibinfo{author}{\bibfnamefont{R.}~\bibnamefont{Eder}},
  \bibinfo{author}{\bibfnamefont{A.-A.} \bibnamefont{Haghighirad}},
  \bibnamefont{et~al.}, \bibinfo{journal}{Phys. Rev. B}
  \textbf{\bibinfo{volume}{99}}, \bibinfo{pages}{035157}
  (\bibinfo{year}{2019}),
  \urlprefix\url{https://link.aps.org/doi/10.1103/PhysRevB.99.035157}.

\bibitem[{\citenamefont{Kushnirenko et~al.}(2020)\citenamefont{Kushnirenko,
  Evtushinsky, Kim, Morozov, Harnagea, Wurmehl, Aswartham, B\"uchner, Chubukov,
  and Borisenko}}]{kush2020}
\bibinfo{author}{\bibfnamefont{Y.~S.} \bibnamefont{Kushnirenko}},
  \bibinfo{author}{\bibfnamefont{D.~V.} \bibnamefont{Evtushinsky}},
  \bibinfo{author}{\bibfnamefont{T.~K.} \bibnamefont{Kim}},
  \bibinfo{author}{\bibfnamefont{I.}~\bibnamefont{Morozov}},
  \bibinfo{author}{\bibfnamefont{L.}~\bibnamefont{Harnagea}},
  \bibinfo{author}{\bibfnamefont{S.}~\bibnamefont{Wurmehl}},
  \bibinfo{author}{\bibfnamefont{S.}~\bibnamefont{Aswartham}},
  \bibinfo{author}{\bibfnamefont{B.}~\bibnamefont{B\"uchner}},
  \bibinfo{author}{\bibfnamefont{A.~V.} \bibnamefont{Chubukov}},
  \bibnamefont{and} \bibinfo{author}{\bibfnamefont{S.~V.}
  \bibnamefont{Borisenko}}, \bibinfo{journal}{Phys. Rev. B}
  \textbf{\bibinfo{volume}{102}}, \bibinfo{pages}{184502}
  (\bibinfo{year}{2020}),
  \urlprefix\url{https://link.aps.org/doi/10.1103/PhysRevB.102.184502}.

\bibitem[{\citenamefont{Dong et~al.}(2009)\citenamefont{Dong, Guan, Zhou, Qiu,
  Ding, Zhang, Patel, Xiao, and Li}}]{dong2009}
\bibinfo{author}{\bibfnamefont{J.~K.} \bibnamefont{Dong}},
  \bibinfo{author}{\bibfnamefont{T.~Y.} \bibnamefont{Guan}},
  \bibinfo{author}{\bibfnamefont{S.~Y.} \bibnamefont{Zhou}},
  \bibinfo{author}{\bibfnamefont{X.}~\bibnamefont{Qiu}},
  \bibinfo{author}{\bibfnamefont{L.}~\bibnamefont{Ding}},
  \bibinfo{author}{\bibfnamefont{C.}~\bibnamefont{Zhang}},
  \bibinfo{author}{\bibfnamefont{U.}~\bibnamefont{Patel}},
  \bibinfo{author}{\bibfnamefont{Z.~L.} \bibnamefont{Xiao}}, \bibnamefont{and}
  \bibinfo{author}{\bibfnamefont{S.~Y.} \bibnamefont{Li}},
  \bibinfo{journal}{Phys. Rev. B} \textbf{\bibinfo{volume}{80}},
  \bibinfo{pages}{024518} (\bibinfo{year}{2009}),
  \urlprefix\url{https://link.aps.org/doi/10.1103/PhysRevB.80.024518}.

\bibitem[{\citenamefont{Bourgeois-Hope
  et~al.}(2016)\citenamefont{Bourgeois-Hope, Chi, Bonn, Liang, Hardy, Wolf,
  Meingast, Doiron-Leyraud, and Taillefer}}]{hope2016}
\bibinfo{author}{\bibfnamefont{P.}~\bibnamefont{Bourgeois-Hope}},
  \bibinfo{author}{\bibfnamefont{S.}~\bibnamefont{Chi}},
  \bibinfo{author}{\bibfnamefont{D.~A.} \bibnamefont{Bonn}},
  \bibinfo{author}{\bibfnamefont{R.}~\bibnamefont{Liang}},
  \bibinfo{author}{\bibfnamefont{W.~N.} \bibnamefont{Hardy}},
  \bibinfo{author}{\bibfnamefont{T.}~\bibnamefont{Wolf}},
  \bibinfo{author}{\bibfnamefont{C.}~\bibnamefont{Meingast}},
  \bibinfo{author}{\bibfnamefont{N.}~\bibnamefont{Doiron-Leyraud}},
  \bibnamefont{and}
  \bibinfo{author}{\bibfnamefont{L.}~\bibnamefont{Taillefer}},
  \bibinfo{journal}{Phys. Rev. Lett.} \textbf{\bibinfo{volume}{117}},
  \bibinfo{pages}{097003} (\bibinfo{year}{2016}),
  \urlprefix\url{https://link.aps.org/doi/10.1103/PhysRevLett.117.097003}.

\bibitem[{\citenamefont{Borzi et~al.}(2007)\citenamefont{Borzi, Grigera,
  Farrell, Perry, Lister, Lee, Tennant, Maeno, and Mackenzie}}]{borzi2007}
\bibinfo{author}{\bibfnamefont{R.~A.} \bibnamefont{Borzi}},
  \bibinfo{author}{\bibfnamefont{S.~A.} \bibnamefont{Grigera}},
  \bibinfo{author}{\bibfnamefont{J.}~\bibnamefont{Farrell}},
  \bibinfo{author}{\bibfnamefont{R.~S.} \bibnamefont{Perry}},
  \bibinfo{author}{\bibfnamefont{S.~J.~S.} \bibnamefont{Lister}},
  \bibinfo{author}{\bibfnamefont{S.~L.} \bibnamefont{Lee}},
  \bibinfo{author}{\bibfnamefont{D.~A.} \bibnamefont{Tennant}},
  \bibinfo{author}{\bibfnamefont{Y.}~\bibnamefont{Maeno}}, \bibnamefont{and}
  \bibinfo{author}{\bibfnamefont{A.~P.} \bibnamefont{Mackenzie}},
  \bibinfo{journal}{Science} \textbf{\bibinfo{volume}{315}},
  \bibinfo{pages}{214} (\bibinfo{year}{2007}),
  \eprint{https://www.science.org/doi/pdf/10.1126/science.1134796},
  \urlprefix\url{https://www.science.org/doi/abs/10.1126/science.1134796}.

\bibitem[{\citenamefont{Kim et~al.}(2008)\citenamefont{Kim, Lawler, Oreto,
  Sachdev, Fradkin, and Kivelson}}]{kim2008}
\bibinfo{author}{\bibfnamefont{E.-A.} \bibnamefont{Kim}},
  \bibinfo{author}{\bibfnamefont{M.~J.} \bibnamefont{Lawler}},
  \bibinfo{author}{\bibfnamefont{P.}~\bibnamefont{Oreto}},
  \bibinfo{author}{\bibfnamefont{S.}~\bibnamefont{Sachdev}},
  \bibinfo{author}{\bibfnamefont{E.}~\bibnamefont{Fradkin}}, \bibnamefont{and}
  \bibinfo{author}{\bibfnamefont{S.~A.} \bibnamefont{Kivelson}},
  \bibinfo{journal}{Phys. Rev. B} \textbf{\bibinfo{volume}{77}},
  \bibinfo{pages}{184514} (\bibinfo{year}{2008}),
  \urlprefix\url{https://link.aps.org/doi/10.1103/PhysRevB.77.184514}.

\bibitem[{\citenamefont{Yamase et~al.}(2005)\citenamefont{Yamase, Oganesyan,
  and Metzner}}]{yamase2005}
\bibinfo{author}{\bibfnamefont{H.}~\bibnamefont{Yamase}},
  \bibinfo{author}{\bibfnamefont{V.}~\bibnamefont{Oganesyan}},
  \bibnamefont{and} \bibinfo{author}{\bibfnamefont{W.}~\bibnamefont{Metzner}},
  \bibinfo{journal}{Phys. Rev. B} \textbf{\bibinfo{volume}{72}},
  \bibinfo{pages}{035114} (\bibinfo{year}{2005}),
  \urlprefix\url{https://link.aps.org/doi/10.1103/PhysRevB.72.035114}.

\bibitem[{\citenamefont{Kreisel et~al.}(2021)\citenamefont{Kreisel, Marques,
  Rhodes, Kong, Berlijn, Fittipaldi, Granata, Vecchione, Wahl, and
  Hirschfeld}}]{Kreisel2021}
\bibinfo{author}{\bibfnamefont{A.}~\bibnamefont{Kreisel}},
  \bibinfo{author}{\bibfnamefont{C.~A.} \bibnamefont{Marques}},
  \bibinfo{author}{\bibfnamefont{L.~C.} \bibnamefont{Rhodes}},
  \bibinfo{author}{\bibfnamefont{X.}~\bibnamefont{Kong}},
  \bibinfo{author}{\bibfnamefont{T.}~\bibnamefont{Berlijn}},
  \bibinfo{author}{\bibfnamefont{R.}~\bibnamefont{Fittipaldi}},
  \bibinfo{author}{\bibfnamefont{V.}~\bibnamefont{Granata}},
  \bibinfo{author}{\bibfnamefont{A.}~\bibnamefont{Vecchione}},
  \bibinfo{author}{\bibfnamefont{P.}~\bibnamefont{Wahl}}, \bibnamefont{and}
  \bibinfo{author}{\bibfnamefont{P.~J.} \bibnamefont{Hirschfeld}},
  \bibinfo{journal}{npj Quantum Materials} \textbf{\bibinfo{volume}{6}},
  \bibinfo{pages}{100} (\bibinfo{year}{2021}), ISSN \bibinfo{issn}{2397-4648},
  \urlprefix\url{https://doi.org/10.1038/s41535-021-00401-x}.

\bibitem[{\citenamefont{Geilikman}(1958)}]{gel}
\bibinfo{author}{\bibfnamefont{B.~T.} \bibnamefont{Geilikman}},
  \bibinfo{journal}{JETP (U.S.S.R.)} \textbf{\bibinfo{volume}{7}},
  \bibinfo{pages}{721} (\bibinfo{year}{1958}).

\bibitem[{\citenamefont{Mineev and Samokin}(1998)}]{mineev}
\bibinfo{author}{\bibfnamefont{V.~P.} \bibnamefont{Mineev}} \bibnamefont{and}
  \bibinfo{author}{\bibfnamefont{K.}~\bibnamefont{Samokin}},
  \emph{\bibinfo{title}{Introduction to Unconventional Superconductivity}}
  (\bibinfo{publisher}{Gordon and Breach Science Publishers},
  \bibinfo{address}{Amsterdam}, \bibinfo{year}{1998}).

\bibitem[{\citenamefont{Arfi}(1993)}]{arfi2}
\bibinfo{author}{\bibfnamefont{B.}~\bibnamefont{Arfi}}, \bibinfo{journal}{Phys.
  Rev. B} \textbf{\bibinfo{volume}{47}}, \bibinfo{pages}{523}
  (\bibinfo{year}{1993}),
  \urlprefix\url{https://link.aps.org/doi/10.1103/PhysRevB.47.523}.

\bibitem[{\citenamefont{Choudhury and Vorontsov}(2021)}]{choudhury2021}
\bibinfo{author}{\bibfnamefont{S.~S.} \bibnamefont{Choudhury}}
  \bibnamefont{and} \bibinfo{author}{\bibfnamefont{A.~B.}
  \bibnamefont{Vorontsov}}, \bibinfo{journal}{Phys. Rev. B}
  \textbf{\bibinfo{volume}{103}}, \bibinfo{pages}{104501}
  (\bibinfo{year}{2021}),
  \urlprefix\url{https://link.aps.org/doi/10.1103/PhysRevB.103.104501}.

\bibitem[{\citenamefont{Yu et~al.}(1992)\citenamefont{Yu, Salamon, Lu, and
  Lee}}]{yu1992}
\bibinfo{author}{\bibfnamefont{R.~C.} \bibnamefont{Yu}},
  \bibinfo{author}{\bibfnamefont{M.~B.} \bibnamefont{Salamon}},
  \bibinfo{author}{\bibfnamefont{J.~P.} \bibnamefont{Lu}}, \bibnamefont{and}
  \bibinfo{author}{\bibfnamefont{W.~C.} \bibnamefont{Lee}},
  \bibinfo{journal}{Phys. Rev. Lett.} \textbf{\bibinfo{volume}{69}},
  \bibinfo{pages}{1431} (\bibinfo{year}{1992}),
  \urlprefix\url{https://link.aps.org/doi/10.1103/PhysRevLett.69.1431}.

\bibitem[{\citenamefont{Ginsberg}(1998)}]{uher}
\bibinfo{author}{\bibfnamefont{D.}~\bibnamefont{Ginsberg}},
  \emph{\bibinfo{title}{Physical Properties of High Temperature Superconductors
  III}} (\bibinfo{publisher}{World Scientific}, \bibinfo{year}{1998}),
  chap.~\bibinfo{chapter}{3}, pp. \bibinfo{pages}{159--283},
  \urlprefix\url{https://www.worldscientific.com/doi/abs/10.1142/9789814439688_0003}.

\end{thebibliography}
\bibliographystyle{apsrev.bst}
\end{document}